\documentclass[fleqn,usenatbib]{mnras}

\usepackage{newtxtext,newtxmath}

\usepackage[T1]{fontenc}

\DeclareRobustCommand{\VAN}[3]{#2}
\let\VANthebibliography\thebibliography
\def\thebibliography{\DeclareRobustCommand{\VAN}[3]{##3}\VANthebibliography}

\usepackage{graphicx}	
\usepackage{amsmath}	
\usepackage{xspace}
\usepackage{multirow}
\graphicspath{{./figures/}}

\newcommand{\planet}{WASP-156~b\xspace}

\newcommand{\Teff}{\ensuremath{T_\mathrm{eff}}\xspace}
\newcommand{\vsini}{\ensuremath{v \sin i_\mathrm{\star}}\xspace}

\newcommand{\Mjup}{\ensuremath{\mathrm{M_{J}}}\xspace}

\newcommand{\Mearth}{\ensuremath{\mathrm{M_{\oplus}}}\xspace}
\newcommand{\Rearth}{\ensuremath{\mathrm{R_{\oplus}}}\xspace}

\newcommand{\ms}{\ensuremath{\mathrm{m\,s^{-1}}}\xspace}
\newcommand{\kms}{\ensuremath{\mathrm{km\,s^{-1}}}\xspace}

\newcommand{\tess}{\textit{TESS}\xspace}

\title[The obliquity of WASP-156]{The Neptunian ridge planet WASP-156 b does not have a polar orbit}

\author[M. Lafarga et al.]{
M.~Lafarga,$^{1,2}$\thanks{E-mail: marina.lafarga-magro@warwick.ac.uk}
J.~I.~Espinoza-Retamal,$^{3}$
H.~M.~Cegla,$^{1,2}$\thanks{UKRI Future Leaders Fellow}
G.~Stefánsson,$^{4,5}$
A.~V.~Freckelton,$^{6}$
A.~Mortier,$^{6}$
\newauthor
S.~Gill,$^{1,2}$
E.~Ahrer,$^{7}$
D.~Anderson,$^{8}$
D.~J.~Armstrong,$^{1,2}$
J.~L.~Bean,$^{9}$ 
V.~Bourrier,$^{10}$
M.~Brady,$^{11}$ 
M.~Brogi$^{12,13}$
\newauthor
E.~M.~Bryant,$^{1,2}$
M.~R.~Burleigh,$^{14}$
L.~Doyle,$^{1,2}$
J.~S.~Jenkins,$^{15,16}$
D.~Kasper,$^{9}$ 
V.~Kunovac,$^{1,2}$
X.~Luo,$^{17}$
\newauthor
L.~Mancini,$^{18,19}$ 
M.~Moyano,$^{8}$
S.~Saha,$^{15,16}$
J.~Southworth,$^{20}$
D.~Veras,$^{1,21,2}$
J.~I.~Vines,$^{8}$
P.~J.~Wheatley$^{1,2}$ 
\newauthor
and 
J.~N.~Winn$^{3}$
\\
$^{1}$Department of Physics, University of Warwick, Gibbet Hill Road, Coventry CV4 7AL, UK\\
$^{2}$Centre for Exoplanets and Habitability, University of Warwick, Gibbet Hill Road, Coventry CV4 7AL, UK\\
$^{3}$Department of Astrophysical Sciences, Princeton University, 4 Ivy Lane, Princeton, NJ 08540, USA\\
$^{4}$Astrophysics \& Space Institute, Schmidt Sciences, New York, NY 10011, USA\\
$^{5}$Anton Pannekoek Institute for Astronomy, University of Amsterdam, Science Park 904, 1098 XH Amsterdam, The Netherlands\\
$^{6}$School of Physics \& Astronomy, University of Birmingham, Edgbaston, Birmingham, B15 2TT, UK\\
$^{7}$Max Planck Institute for Astronomy (MPIA), K\"{o}nigstuhl 17, D-69117 Heidelberg, Germany\\
$^{8}$Instituto de Astronom\'ia, Universidad Cat\'olica del Norte, Angamos 0610, 1270709, Antofagasta, Chile\\
$^{9}$Department of Astronomy \& Astrophysics, University of Chicago, Chicago, IL 60637, USA\\
$^{10}$Observatoire Astronomique de l’Université de Genève, Chemin Pegasi 51b, 1290 Versoix, Switzerland\\
$^{11}$Department of Physics \& Astronomy, Michigan State University, East Lansing, MI 48824, USA\\
$^{12}$Department of Physics, University of Torino, Via Pietro Giuria 1, I-10125 Torino, Italy\\
$^{13}$INAF–Osservatorio Astrofisico di Torino, Via Osservatorio 20, I-10025 Pino Torinese, Italy\\
$^{14}$School of Physics \& Astronomy, University of Leicester, Leicester, LE1 7RH, UK\\
$^{15}$Instituto de Estudios Astrof\'isicos, Facultad de Ingenier\'ia y Ciencias, Universidad Diego Portales, Av. Ej\'ercito 441, Santiago, Chile\\
$^{16}$Centro de Astrof\'isica y Tecnolog\'ias Afines (CATA), Casilla 36-D, Santiago, Chile\\
$^{17}$School of Physics and Astronomy, China West Normal University, Nanchong 637009, People's Republic of China\\
$^{18}$Department of Physics, University of Rome ``Tor Vergata'', Via della Ricerca Scientifica 1, 00133 Rome, Italy\\
$^{19}$INAF -- Turin Astrophysical Observatory, via Osservatorio 20, 10025 Pino Torinese, Italy\\
$^{20}$Astrophysics Group, Keele University, Staffordshire ST5 5BG, UK\\
$^{21}$Centre for Space Domain Awareness, University of Warwick, Coventry CV4 7AL, UK\\
}
\date{Accepted 2026 May 26. Received 2026 May 26; in original form 2026 April 22}

\pubyear{2026}

\begin{document}
\label{firstpage}
\pagerange{\pageref{firstpage}--\pageref{lastpage}}
\maketitle

\begin{abstract}
The population of short-period exo-Neptunes is thought to be shaped by an interplay between different dynamical mechanisms, such as orbital migration and tidal effects, and photoevaporation.
We can gain insight into these processes by studying observables such as the stellar obliquity.
Here we study the Rossiter-McLaughlin (RM) effect and measure the projected obliquity, $\lambda$, of the Neptunian ridge planet WASP-156~b.
We analyse new ESPRESSO and MAROON-X spectroscopic transit observations, and new NGTS photometry simultaneous to the ESPRESSO data.
Our analyses show an aligned orbit
($\lambda=-8\pm16\degr$, based on the ESPRESSO observations), 
in contrast to a previous report of a highly misaligned orbit. We also find the star's projected rotational velocity to be $\vsini\leq2$~\kms from spectral line modelling and 
$\vsini=0.40\pm0.11$~\kms from the RM modelling. This is lower than the previously reported value of $\sim4$~\kms, which could partly explain the previously derived polar orbit. 
We also update the system's orbital parameters and rule out Jupiter-mass companions within 5~au using long-term radial velocity data.
The planet's aligned and circular orbit ($e<0.16$ at $3\sigma$), and lack of nearby massive companions, are consistent with in situ formation or early disc-driven migration.
Our findings move WASP-156~b from a tentative cluster of close-in Neptunes in polar orbits to the group of aligned Neptunes.
\end{abstract}




\section{Introduction}

\begin{figure}
    \centering
    \includegraphics[width=\linewidth]{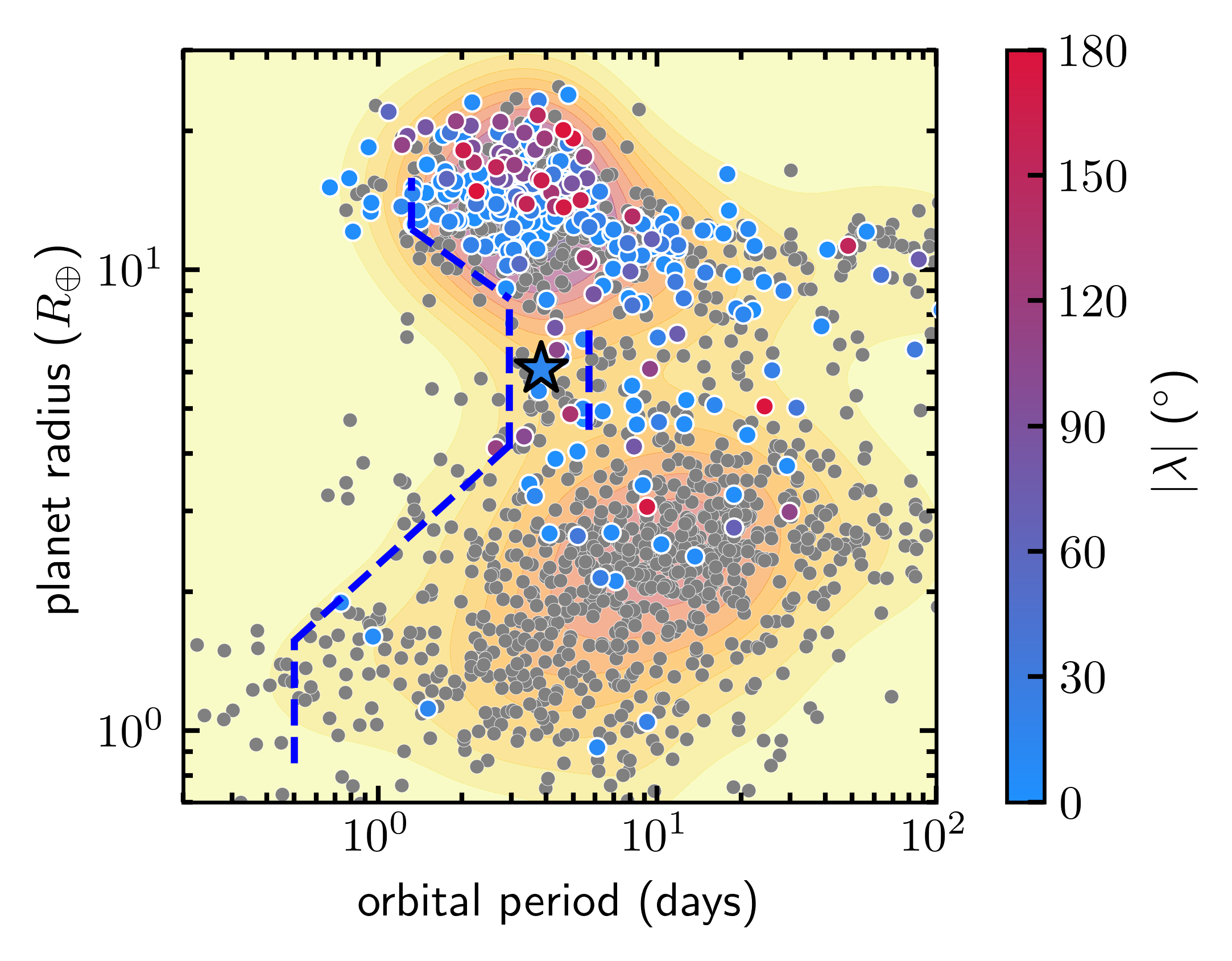}
    \caption{Orbital period versus planet radius diagram. Grey points are transiting exoplanets from the TEPCat catalogue \citep[][accessed on 2026/05/11]{Southworth2011,southworth2026tepcat}. Systems with a measured projected obliquity are coloured according to that value. WASP-156~b is highlighted as a star following the same colour code. The dashed blue lines indicate the boundaries of the Neptune desert, savanna, and ridge from \citet{Castro-Gonzalez2024}. Contours show the density of points.}
    \label{fig:neptune_desert}
\end{figure} 

With over 6000 exoplanets discovered to date,\footnote{NASA Exoplanet Archive \url{https://exoplanetarchive.ipac.caltech.edu/}} studies of planetary demographics are increasingly powerful. Short-period Neptunes (orbital period $P\lesssim30$~d, planetary radius $2 \lesssim R_\mathrm{p} \lesssim 8\,\Rearth$) are an interesting population showing a diverse landscape that traces a variety of formation and evolution processes. A combination of orbital migration, tidal decay, and photoevaporation \citep[e.g.][]{Matsakos2016,davis2009evaporation,Owen2013,Ionov2018,owen2018desert} is thought to be responsible for the observed scarcity of very close-in ($P\lesssim3$~d) Neptunes, the so-called hot Neptunian `desert' \citep[e.g.][see Figure \ref{fig:neptune_desert}]{lecavelierdesetangs2007evaporation,szabo2011desert,lundkvist2016desert,mazeh2016desert,Castro-Gonzalez2024,Cui2026}. The occurrence of Neptunes mildly increases at longer periods ($P\gtrsim5$~d) in a Neptunian `savanna', implying different formation and evolution mechanisms for the two regions \citep{bourrier2023dream}. A Neptunian `ridge' between the desert and the savanna has recently been proposed from an observed overabundance of the \textit{Kepler} sample with periods between around 3 and 6~d \citep{Castro-Gonzalez2024}.

The host star's obliquity, $\psi$, the angle between its spin axis and a planet's orbital axis (or $\lambda$, its sky-projection), is a useful observable that provides information on the planet's formation and evolution. 
The most simplistic picture assumes that protoplanetary discs are aligned with the stellar spin, since both originate from the same rotating molecular cloud core, resulting in stars with primordially low obliquities. Aligned close-in planets are then thought to have migrated inwards early-on via disc driven migration \citep[e.g.][]{Goldreich1979,Lin1996}, while misaligned close-in planets should have experienced high-eccentricity tidal migration due to gravitational interactions with another body capable of altering the primordial aligned obliquity, such as planet-planet scattering \citep[e.g.][]{Ford2008,Chatterjee2008,Nagasawa2011,Beauge2012}, secular cyclic interactions \citep[e.g.][]{vonZeipel1910,Kozai1962,Lidov1962,WuMurray2003,Naoz2011,Teyssandier2013,Naoz2016}, or secular chaos \citep[e.g.][]{WuLithwick2011,Hamers2017}.
However, reality might be more complex. Primordial misalignments are also possible \citep[e.g.][]{Batygin2012,Bate2018,Biddle2025}, and interactions with the disc and tidal dissipation can result in misaligned obliquities being damped into aligned orbits \citep[e.g.][]{Fabrycky2007,Nagasawa2008,Winn2010,Correia2011,Albrecht2012,Teyssandier2019}.

A wide range of obliquities have been observed, including well-aligned planets, planets in polar orbits, and retrograde planets \citep[see e.g.][]{Albrecht2022,Mancini2022}.
Observational data seem to show a tentative dichotomy in the obliquities of giant planets and Neptunes, with clusters at 0\degr\ and 90\degr\ \citep[e.g.][]{albrecht2021preponderance,bourrier2023dream}. Tentative empirical evidence of a significant preponderance of polar orbits among Neptunes has been found using statistical analyses \citep{Espinoza-Retamal2024,Knudstrup2024,Rossi2026}. However, the number of Neptunes with measured obliquities is still small, and a larger sample of planets is needed to reach a robust conclusion.

One of the most direct and accurate ways to estimate obliquities is through the well-known Rossiter-McLaughlin effect \citep[RM;][]{Rossiter1924,McLaughlin1924,Albrecht2022}.
As a planet transits its host star, it blocks different parts of the rotating stellar disc, which results in a distortion of the observed stellar spectral lines. This affects the disc-integrated radial velocities (RVs) measured from the stellar spectra in a manner that depends on the sky-projected obliquity $\lambda$ and the projected rotational velocity of the star \vsini \citep[see, e.g.][]{Triaud2018}.
The amplitude of the RM distortion depends on the area of the stellar disc occulted by the planet and the rotation of the star. Hence, the method is more sensitive to large, close-in planets and stars that are fast rotators, which yield RM signals with larger amplitudes, easier to detect.

In this work, we study the obliquity of the Neptunian-ridge planet \planet.
\planet is a hot super-Neptune ($R_\mathrm{p}\simeq6.1\,\Rearth$, $M_\mathrm{p}\simeq42\,\Mearth$ and $P\simeq3.84$~d; see Figure~\ref{fig:neptune_desert}) orbiting a relatively bright ($V=11.6$~mag) K3~V star \citep[][also see Table~\ref{tab:juliet_params} for the system parameters used in this work]{demangeon2018wasp}.

Here, we report on two new spectroscopic transit observations obtained with the Echelle SPectrograph for Rocky Exoplanets and Stable Spectroscopic Observations \citep[ESPRESSO;][]{pepe2021espresso} and one new transit observed with the M-dwarf Advanced Radial velocity Observer Of Neighboring eXoplanets \citep[MAROON-X;][]{seifahrt2016maroonx,seifahrt2018maroonx}, and use them to measure the obliquity of the system.
Previously, \citet{bourrier2023dream} had reported an RM measurement of the star's projected obliquity that indicated a nearly polar orbit, based on observations of one transit with the visible arm of the Calar Alto high-Resolution search for M dwarfs with Exoearths with Near-infrared and optical Échelle Spectrographs \citep[CARMENES; mounted at the 3.5~m telescope in Calar Alto, Spain, with wavelength range $5200-9600$~\AA\ and resolving power $R=94\,600$;][]{quirrenbach2014carmenes,quirrenbach2018carmenes}.
Recently, \citet{Jiang2026} used the first night of ESPRESSO observations and derived an obliquity compatible with an aligned orbit, contradicting the CARMENES results.
Compared to CARMENES, both ESPRESSO and MAROON-X are mounted at larger telescopes, the European Southern Observatory's (ESO) 8.2~m Very Large Telescope (VLT) in Paranal, Chile, and the 8.1~m Gemini North telescope in Mauna Kea, Hawai'i, USA, respectively. These telescopes have a larger collecting area than the 3.5\,m Calar Alto telescope, allowing for a higher signal-to-noise ratio (S/N), shorter exposure times, and better time sampling of the transit.

In Sect.~\ref{sec:obs}, we present the observations used in this work. We derive updated spectroscopic stellar parameters of the host star in Sect.~\ref{sec:stellarparams} and, in Sect.~\ref{sec:refit}, we perform a re-characterisation of the star-planet system to obtain updated parameters. In Sect.~\ref{sec:rm}, we present our spectroscopic transit analyses exploiting the RM effect. We discuss the previous and new obliquity measurements of \planet, explore its possible formation mechanisms, and put the system in context in Sect. \ref{sec:discussion}. Finally, we conclude our work in Sect. \ref{sec:conclusion}.


\section{Observations}\label{sec:obs}

\begin{table}
\centering
\caption{Spectroscopic observations and average S/N.}
\begin{tabular}{cccc}
\hline
Instrument & Night & \# observations & Median S/N per pixel \\ \hline
\multirow{2}{*}{ESPRESSO}  & 2022/09/02 & 21 (11/10) & 85 \\
                           & 2022/09/29 & 21 (10/11) & 70 \\
MAROON-X                   & 2021/11/04 & 38 (17/21) & 71 \\
\hline
\end{tabular}
\begin{flushleft}
{\it Notes:} \# observations shows the total number of observations, and inside the brackets is the number of in-transit / out-of-transit observations. The per pixel median S/N is that of order 87 (central wavelength $\sim$703~nm), one of the orders with the highest S/N for both spectrographs for WASP-156. Note that the S/N that we quote here corresponds to the values of the individual slices into which order 87 is divided added in quadrature (two slices for ESPRESSO and three slices for MAROON-X).
\end{flushleft}
\label{tab:obs}
\end{table}

\subsection{Spectroscopic data}

We observed two transits of \planet with ESPRESSO and one with MAROON-X. We show a summary of the observations in Table~\ref{tab:obs} and Figure~\ref{fig:obs}, and detail them in the following sub-sections.

\subsubsection{ESPRESSO transits}

We observed two transits of \planet with the high-resolution, optical spectrograph ESPRESSO \citep[wavelength range 3782--7887~\AA{};][]{pepe2021espresso} installed on the VLT at ESO's Paranal Observatory, in Chile (ESO programme ID: 109.23FU, PI: Lafarga). The observations were performed in 1-UT configuration (on UT 2, Kueyen) and high-resolution mode HR21 (i.e., with $2\times1$ readout binning, median resolving power of $R=138\,000$, and spectral sampling of 4.5 pixels per spectral element). The target star was observed with fibre A. Fibre B was used to simultaneously monitor the sky. We obtained the raw data from the ESO archive and reduced them with the ESPRESSO Data Reduction Software\footnote{\url{https://www.eso.org/sci/software/pipelines/espresso/espresso-pipe-recipes.html}} (DRS) version 3.0.0, which performs standard reduction steps for \'echelle spectra \citep[for details, see][]{pepe2021espresso}. In the following analysis, we used the blaze-corrected and sky-subtracted (i.e., corrected for telluric sky emission) 2D spectra.

During the first transit (2022/09/02), because the measured flux in the first exposure was lower than expected, we decided to increase the exposure time from 600 to 700\,s for subsequent exposures. During the second transit (2022/09/29), we kept the exposure time at 600\,s since the flux difference between 600 and 700\,s did not seem to affect our preliminary results (i.e., the RV uncertainty difference between the two different exposure times was small). Close to the end of the second transit, observations had to be aborted because of pointing restrictions due to wind. Observations restarted after the end of the transit, missing the transit egress.
The observations of the first night covered airmasses from $\sim1.5$ to $1$ and a line-of-sight seeing below $1''$ for most of the observations, while the second night covered airmasses from $\sim2$ to $1$, with the seeing close to $1''$. For both nights, most in-transit observations occurred at airmasses below $1.5$.
As a result of the slightly longer exposure times, observations of the first night reached per pixel signal-to-noise ratios close to 85 at 703~nm (order 87, where the spectrograph achieves a high S/N for this target), while for the second night, the S/N at that same wavelength was close to 70. Note that this is the S/N that results from adding in quadrature the S/N of the two slices into which order 87 is divided for ESPRESSO.

\subsubsection{MAROON-X transit}

We also observed one transit of \planet with MAROON-X \citep[][]{seifahrt2016maroonx,seifahrt2018maroonx,seifhart2020} at the 8.1~m Gemini North telescope in Hawai'i on 2021/11/04 (programme ID: GN-2021B-Q-218, PI: Stefánsson). MAROON-X is a high-resolution ($R=85\,000$, spectral sampling of 3.5 pixels per spectral element), optical-red spectrograph with a blue and a red channel, covering the wavelength ranges 5000--6700~\AA{} and 6500--9200~\AA{}. The data was reduced with a custom Python pipeline that produces extracted 2D spectra.

The MAROON-X observations were performed with an exposure time of 400~s and cover airmasses from $\sim$3 before the start of the transit down to $\sim$1 towards the end of the transit and later out-of-transit observations. The in-transit observations in particular cover airmasses from $\sim$2 to $\sim$1. 
The seeing during the pre-out-of-transit observations was of $\sim1.5''$, and it decreased down to $\sim0.75''$ close to the end of the transit. The post-out-of-transit observations (close to phase 0.15) reached a seeing of $\sim0.5-0.6''$.
The S/N per pixel is close to 60 at $\sim703$~nm at the start of the observations and increases to 80--100 towards the end of the transit, likely reflecting the decrease in airmass and seeing. Again, note that the S/N values quoted here are those corresponding to the three slices that order 87 is divided into for MAROON-X.

\subsubsection{Archival out-of-transit data}

WASP-156 has previously been observed with the spectrographs CORALIE \citep{Queloz2001}, Spectrographe pour l’Observation des Phénomènes des Intérieurs stellaires et des Exoplanètes \citep[SOPHIE;][]{Perruchot2008}, High Resolution Echelle Spectrometer \citep[HIRES;][]{Vogt1994}, and High Accuracy Radial velocity Planet Searcher \citep[HARPS;][]{Mayor2003} in out-of-transit phases. The CORALIE and SOPHIE data were used to initially confirm the nature of the planet and measure its mass \citep{demangeon2018wasp}. HIRES data were used by \citet{Polanski2024} to refine the parameters of the planet and perform a uniform analysis together with other targets of the \tess-Keck survey. The HARPS data has never been published in the past and is publicly available in the ESO Archive.\footnote{\url{https://archive.eso.org/eso/eso_archive_main.html}, Programme 0102.C-0618(A)} These archival observations of WASP-156, along with the best-fitting model, are shown in Figure~\ref{fig:juliet_fit}.

We note that there exist CARMENES \citep[see][]{bourrier2023dream} and HARPS spectra obtained in transit for RM measurements. We decided not to include these data here because they have significantly lower S/N than the new ESPRESSO and MAROON-X data we obtained. This is mainly due to the larger collecting power of the VLT and Gemini North telescopes. We do not expect the derived parameters to improve by adding lower S/N CARMENES and HARPS data; therefore, we restrict our spectroscopic transit analysis to the ESPRESSO and MAROON-X data.


\subsection{Photometric data}

\subsubsection{TESS}

To better constrain the parameters of the planet and its orbit, in this work, we also made use of \tess \citep[Transiting Exoplanet Survey Satellite;][]{Ricker2015} photometric data. 
For the past eight years, \tess has been observing nearly the whole sky in the form of 27-day sectors. WASP-156 has been observed by \tess in sectors 4, 31, 42, 43, 70, and 71. None of the sectors coincides with our ESPRESSO or MAROON-X spectroscopic transit observations. In particular, we worked with the combined 2-min light curves from these sectors. These light curves were processed using the Science Processing Operations Center (SPOC) pipeline \citep{spoc}, and are corrected for pointing- or focus-related instrumental signatures, discontinuities resulting from radiation events in the CCD detectors, outliers, and flux contamination from nearby sources. We searched for, downloaded, and combined all the light curves using the \texttt{lightkurve} package \citep{Lightkurve}. The phase-folded \tess light curve, along with the best-fitting model (see Sect. \ref{sec:refit}), is shown in Figure~\ref{fig:juliet_fit}.


\subsubsection{NGTS}

We used the Next-Generation Transit Survey \citep[NGTS;][]{Wheatley2018} to observe the transits of \planet simultaneously with the ESPRESSO observations (i.e., on 2022/09/02 and 2022/09/29). The NGTS facility consists of twelve telescopes with 20~cm diameter and is located in ESO's Paranal Observatory, Chile. The NGTS cameras are independent, have a custom optical-red filter covering the wavelength range 520--890~nm, a wide field-of-view of $2.8 \times 2.8$~\degr{} each, and a pixel scale of 5~arcsec. By observing the same star simultaneously with several cameras, NGTS achieves high-precision photometry of exoplanet transits \citep{Smith2020,bryant2020}.

WASP-156 was observed in 10~s exposure times with four cameras on the first night of ESPRESSO observations (2022/09/02), and with five cameras on the second night (2022/09/29). The observations were reduced with a custom aperture photometry pipeline that utilises the SEP library \citep{Bertin1996, Barbary2016} for source extraction and photometry. We used an aperture radius of 3~pixels for all cameras and nights. A time series of differential flux is produced using a set of unblended comparison stars similar in apparent magnitude, colour, and nearby position in the CCD to WASP-156, selected using \textit{Gaia} information \citep{gaia2018dr2,gaia2023dr3}.


\section{Stellar parameters}\label{sec:stellarparams}

\subsection{Stellar parameters from spectroscopy}\label{sec:paws}

We derived stellar parameters (effective temperature $T_\mathrm{eff}$, surface gravity $\log g$, metallicity [Fe/H], and projected equatorial rotational velocity \vsini) from the ESPRESSO observations using the \texttt{PAWS}\footnote{\url{https://github.com/alixviolet/PAWS}} pipeline \citep{freckelton2024paws}.
The curve-of-growth equivalent widths method was used to generate initial estimates of the stellar parameters. We employed the WIDTH \citep{sbordone2004} radiative transfer code and solar input parameters. Spectral synthesis, using the SPECTRUM \citep{gray1994} radiative transfer code, was then used to refine these estimates, in addition to determining the macroturbulence velocity and projected rotational velocity. As is standard in \texttt{PAWS}, both the SPECTRUM line list \citep{gray1994} and ATLAS model atmospheres \citep{kurucz2005} were used by both methods. 

We ran \texttt{PAWS} on four different sets of 1D co-added rest-frame-shifted spectra: 1) all the observations of the first night, 2) the first 15 observations of the second night, 3) the last 6 observations of the second night, and 4) all observations co-added.
Sets 1 to 3 were obtained directly from the ESPRESSO DRS pipeline, and we used \texttt{PAWS} to co-add all the observations for set 4. The ESPRESSO DRS provided two co-added spectra for the second night instead of one because the observations were aborted (due to pointing restrictions) and restarted towards the end of the transit, resulting in two different observing blocks.

Table~\ref{tab:paws} shows the stellar parameters obtained by \texttt{PAWS} on the ESPRESSO data, for the four different sets of co-added spectra considered. All sets show extremely similar results, showcasing the quality of the ESPRESSO data. 
The $T_\mathrm{eff}$, $\log g$, and [Fe/H] are in agreement (within 1~$\sigma$) with previous literature results obtained in \citet[][$T_\mathrm{eff}=4910\pm$61~K, $\log g=4.60^{+0.04}_{-0.07}$, $\mathrm{[Fe/H]}=0.24\pm0.12$]{demangeon2018wasp} from 40 co-added SOPHIE spectra \citep[$R$\,=\,40\,000,][]{bouchy2009sophie} using methods from \citet{doyle2013wasp}, and from Gaia photometry \citep[][$T_\mathrm{eff}=4970^{+49}_{-118}$~K]{gaia2018dr2}.

Regarding the \vsini, \texttt{PAWS} converged on a value close to $\sim$0.9~\kms. However, below 2~\kms, it is challenging to accurately distinguish between different broadening mechanisms (e.g., instrumental, rotational, macroturbulence). Therefore, we constrain the \vsini to an upper limit of 2~\kms.
This value is lower than the literature value of 3.8$\pm$0.91~\kms from \citet{demangeon2018wasp}, obtained from the SOPHIE spectra by fitting the profiles of several unblended lines, which, as reported, also agrees with a method that uses the cross-correlation function (CCF) profiles following \citet{boisse2010sophieJupiter}. 
Note that both methods are based on empirically calibrated relations. The line-fitting method used a macroturbulence value based on the calibration of \citet{Doyle2014}, valid for $T_\mathrm{eff}$ between 5200 and 6400~K, which had to be extrapolated for WASP-156. The method using CCFs is based on an empirically calibrated relation obtained for SOPHIE, based on a relation between the SOPHIE CCF widths and the star's \vsini computed for a range of stars with known \vsini. Hence, these methods are sensitive to the stars used in the calibration. As explained above, \texttt{PAWS} uses a different methodology based on spectral synthesis, which could explain the difference in \vsini. An in-depth analysis on this \vsini discrepancy is out of the scope of this work, however, for the remainder of this article, we consider the upper limit of 2~\kms obtained with \texttt{PAWS} from ESPRESSO, and not the previous SOPHIE-based value, due to the higher resolution of ESPRESSO and the \texttt{PAWS} methodology, which does not rely on calibrations.

\begin{table}
\centering
\caption{Spectroscopic stellar parameters derived with \texttt{PAWS}.}
\begin{tabular}{lcccc}
\hline
Observation    & $T_\mathrm{eff}$ & $\log g$     & [Fe/H]        & \vsini  \\ 
set            & (K)              &              &               & (\kms)      \\ \hline
1) 2022/09/02  & 5022$\pm$105     & 4.47$\pm$0.10 & 0.31$\pm$0.15 & $\leq$2.0   \\
2) 2022/09/29 1& 5022$\pm$105     & 4.47$\pm$0.10 & 0.31$\pm$0.15 & $\leq$2.0   \\
3) 2022/09/29 2& 5022$\pm$105     & 4.47$\pm$0.10 & 0.31$\pm$0.15 & $\leq$2.0   \\
4) All         & 5036$\pm$104     & 4.46$\pm$0.08& 0.32$\pm$0.15 & $\leq$2.0   \\
\hline
\end{tabular}
\begin{flushleft}
\end{flushleft}
\label{tab:paws}
\end{table}

\subsection{Stellar rotation period}\label{sec:prot}

The stellar rotation period of the star is currently unknown. Unless the star has a low stellar inclination (i.e., it is seen close to pole-on), the low value of \vsini would suggest a long stellar rotation period.
The stellar rotation period can be used to disentangle the stellar inclination from \vsini, which allows us to measure the true 3D obliquity instead of the projected one. 

We performed an archival search of long-term photometry to try to estimate the stellar rotation period of WASP-156. We looked for periodicities in the All-Sky Automated Survey for Supernovae \citep[ASAS-SN][]{Shappee2014,Kochanek2017}\footnote{\url{https://asas-sn.osu.edu/}} and the Hungarian-made Automated Telescope PI Steradians (HATPI) data.\footnote{\url{https://hatpi.org}} Our analysis did not result in a clear signal that we can confidently assign to the stellar rotation, see details in Appendix \ref{sec:prot_appendix}.
We also checked the \tess All-Sky Rotation Survey (TARS) catalogue \citep{Boyle2026}, a recent rotation survey based on \tess data, but WASP-156 does not appear in it, implying that a rotation-related signal has not been detected for this star in \tess.


\section{Re-characterisation of the system} \label{sec:refit}

To refine the parameters of WASP-156~b, we performed a joint analysis of all the available \tess and NGTS photometry and out-of-transit CORALIE, SOPHIE, HARPS, and HIRES RVs using the \texttt{juliet} \citep{juliet} code. \texttt{juliet} uses \texttt{batman} \citep{batman} for the photometric transit model, \texttt{radvel} \citep{radvel} for the Keplerian orbit model, and the \texttt{dynesty} dynamic nested sampler \citep{dynesty} to obtain posterior probability distributions. For the fit, we adopted 4500 live points, and the default configurations of the sampler for the bounding and sampling methods (i.e., multi-ellipsoidal bounding and random-walk sampling). The runs were stopped when the change in the logarithm of the Bayesian evidence ($\log{Z}$) was less than 0.01, which is the default convergence criterion. We placed uninformative priors on the parameters, with an informative Gaussian prior on the stellar density of $\rho_\star=2.154\pm150$ g cm$^{-3}$ based on the value reported by \citet{MacDougall2023}. Limb darkening parameters were sampled using the quadratic $q_1$ and $q_2$ parametrization from \citet{Kipping13} with uniform priors. To model the RVs we placed an uninformative prior on the RV semi-amplitude $K$. We included independent RV offsets ($\gamma_{\rm inst}$) and log-uniform jitter terms ($\sigma_{\rm inst}$) for each instrument to account for possible systematics.
For the NGTS transits, we treated each camera separately with shared limb-darkening coefficients, and detrended the light curves with airmass during the fit.
Finally, to account for variability and systematic noise in the \tess light curve, we included a Matern-3/2 Gaussian Process as implemented in \texttt{celerite} \citep{celerite} and available in \texttt{juliet}. 

\begin{figure*}
    \centering
    \includegraphics[width=0.49\textwidth]{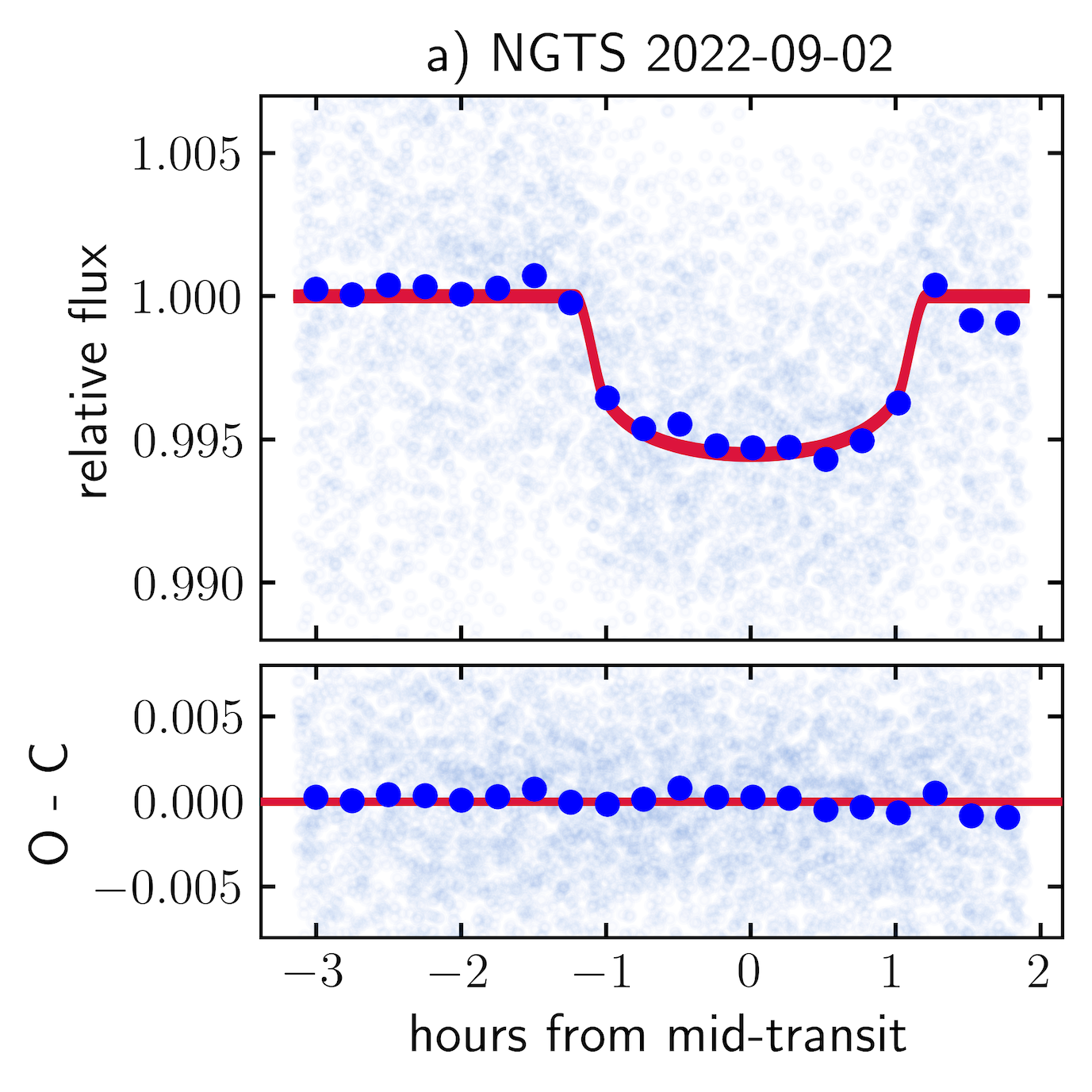}
    \includegraphics[width=0.49\textwidth]{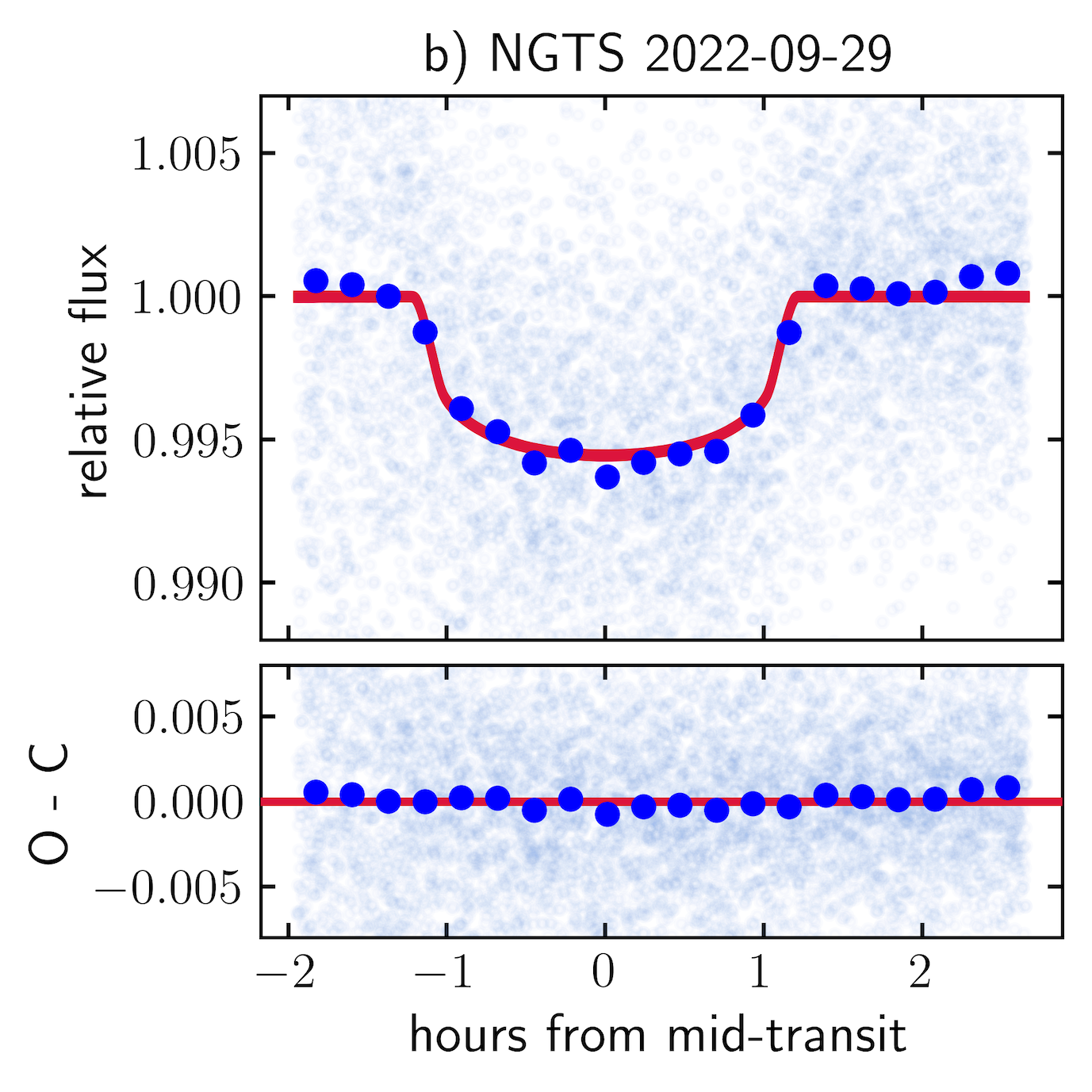}\\
    \includegraphics[width=0.49\textwidth]{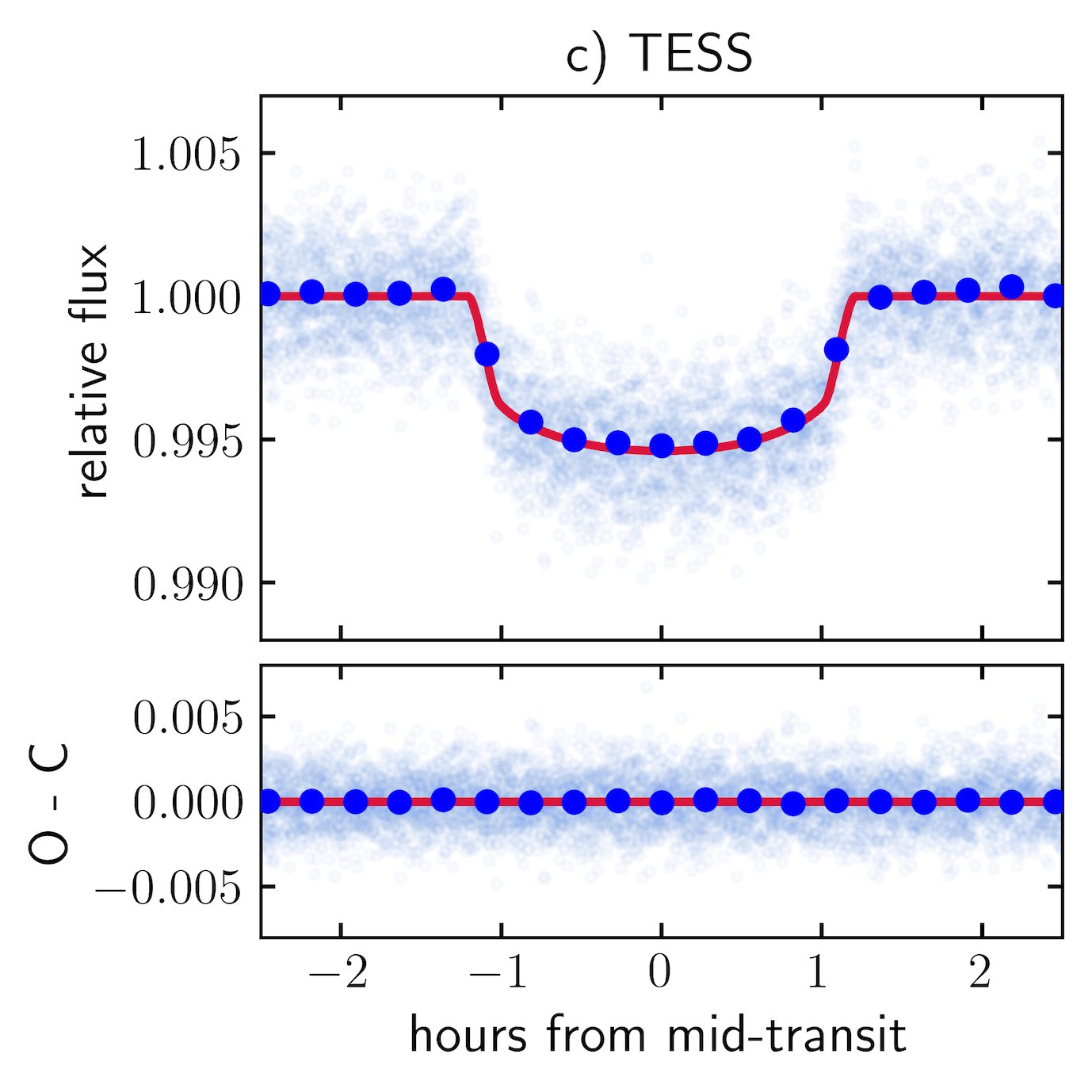}
    \includegraphics[width=0.49\textwidth]{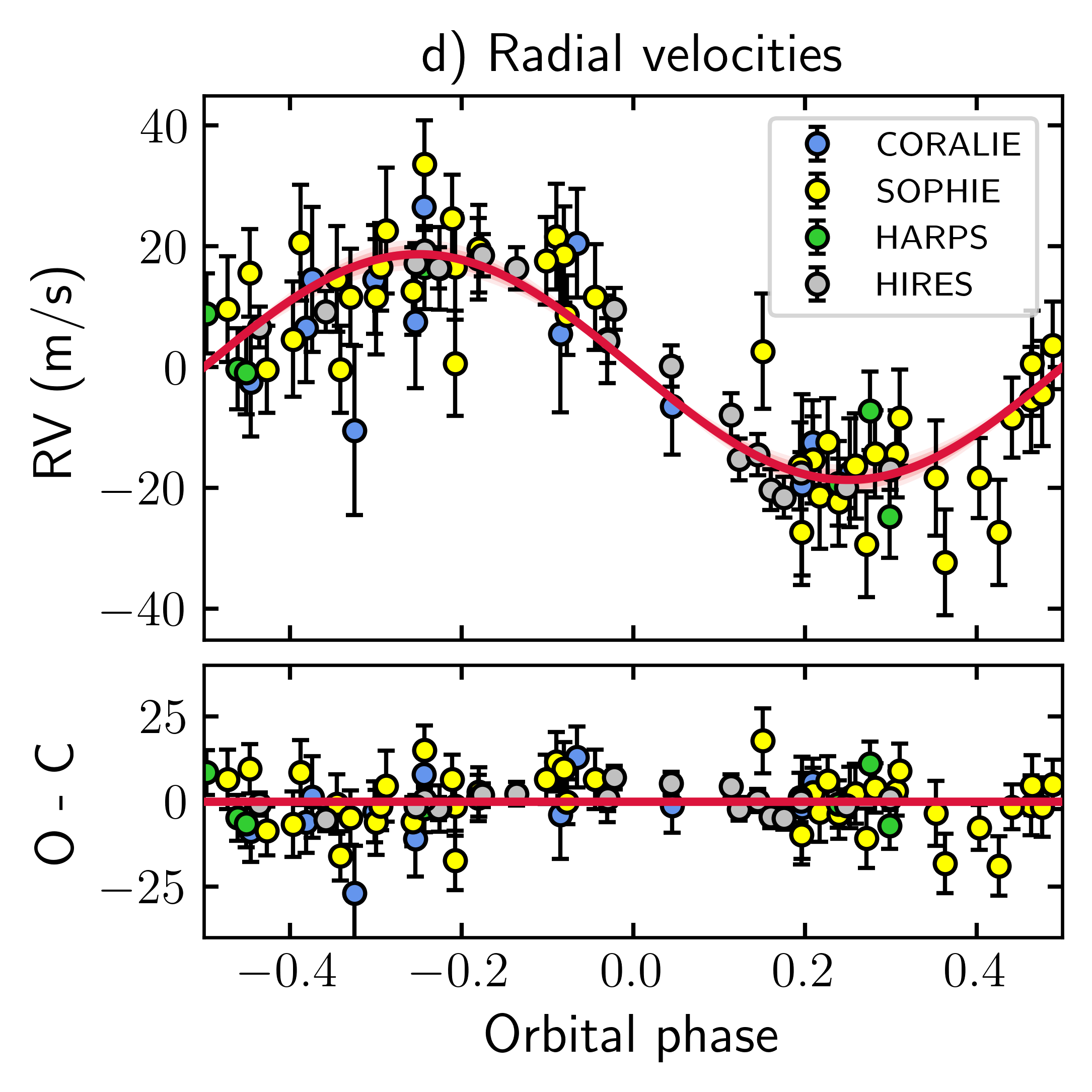}
\caption{\tess, NGTS, and archival RV observations of WASP-156 along with the best-fitting model from \texttt{juliet}. a) and b) airmass-detrended NGTS transits from 2022/09/02 and 2022/09/29, respectively, where large filled circles show the binned data of all cameras on a timescale of about 15 minutes. c) Phase-folded and GP-detrended \tess light curve. 
d) Phase-folded out-of-transit radial velocities taken with different instruments. Shaded red areas represent $1\sigma$, $2\sigma$, and $3\sigma$ models (the confidence intervals are too small to be seen clearly).}
    \label{fig:juliet_fit}
\end{figure*}

The final parameters derived from this analysis are shown in Tables~\ref{tab:juliet_params} (main orbital parameters) and \ref{tab:juliet_params_extra} (additional instrumental parameters), and the photometric and RV data, along with the best-fitting model, are shown in Figure~\ref{fig:juliet_fit}. Our newly derived parameters generally agree with literature values within $1\,\sigma$ and have comparable or in some cases smaller uncertainties, probably due to the large amount of data used in our fit. In particular, we obtained improved uncertainties on the orbital period $P$ 
compared with \citet[][$6\cdot10^{-7}$ vs $9.7\cdot10^{-7}$~d, about 1.6 times more precise]{Kokori2023}, better planetary-to-stellar radius ratio $R_\mathrm{p}/R_\star$ uncertainties compared with \citet[][$7\cdot10^{-4}$ vs $8.7\cdot10^{-4}$, 1.2 times more precise]{patel2022ldc}, and better RV semi-amplitude $K$ uncertainties compared with \citet[][0.8 vs 1~\ms, 1.25 times more precise]{demangeon2018wasp}. 
Regarding derived parameters, we improved on the planetary mass $M_\mathrm{p}$ and density $\rho_\mathrm{p}$ compared with \citet[][2 vs 2.9~\Mearth, 1.5 times more precise, and 0.07 vs 0.18~g~cm$^{-3}$, 2.6 times more precise; note that for that work we also improved on the reported $K$, 1.1~\ms, which drives the improvement in mass]{Polanski2024}.
Note that here we are directly comparing the uncertainties, since the parameter values are similar.
We consider the literature parameters with smaller uncertainties. These come from different works, therefore, the different values that we are using to compare our own values to are not necessarily self-consistent with each other.

\begin{table}
    \centering
    \caption{Summary statistics of posterior distributions of the main orbital parameters of the \texttt{juliet} fit. The posterior summary statistics for the rest of parameters are shown in Table \ref{tab:juliet_params_extra}.}
    \begin{tabular}{lr}
    \hline
       Sampled Parameter & Posterior \\ \hline
       $P$ (d)  & $3.8361629\pm0.0000006$\\
       $t_0$ (BJD $-$ 2,450,000) & $9058.6113\pm0.0001$\\
       $R_\mathrm{p}/R_{\star}$ & $0.0682\pm0.0007$\\
       $b$ & $0.40\pm0.05$\\
       $\rho_\star$ (g cm$^{-3}$) & $2.25\pm0.14$\\
       $e$ & 0 (fixed, $<0.16$ at $3\sigma$)\\
       $K$ (m s$^{-1}$) & $18.6\pm0.8$\\
    \hline
       $q_1^{\rm TESS}$ & $0.32^{+0.13}_{-0.10}$\\ [2pt] %
       $q_2^{\rm TESS}$ & $0.43^{+0.22}_{-0.17}$\\ [2pt] %
       $q_1^{\rm NGTS}$ & $0.53^{+0.21}_{-0.17}$\\ [2pt] %
       $q_2^{\rm NGTS}$ & $0.31^{+0.25}_{-0.18}$\\ [2pt] %
    \hline
    Derived Parameters & \\
    \hline
    $a/R_{\star}$ & $12.04\pm0.26$\\
    $i$ ($^{\circ}$) & $88.1\pm0.3$\\
    $a$ (au) & $0.046\pm0.001$\\
    $M_\mathrm{p}$ (\Mearth) & $42\pm2$\\
    $R_\mathrm{p}$ (\Rearth) & $6.15\pm0.12$\\
    $\rho_\mathrm{p}$ (g cm$^{-3}$) & $0.990\pm0.070$ \\
    \hline
    \end{tabular}
    \label{tab:juliet_params}
\end{table}


\section{Spectroscopic transit analysis}\label{sec:rm}

To analyse the RM effect in the ESPRESSO and MAROON-X observations, we first computed the standard disc-integrated RVs with two different methods: the CCF and the template matching approaches.
We then used two different techniques that exploit the RM effect to measure the stellar obliquity of the system: the classical RM and the reloaded RM. 
We describe in detail our methodology in the following subsections.

\subsection{Disc-integrated RV computation} \label{sec:rv_computation}

For the ESPRESSO observations, we used the CCFs and disc-integrated RVs computed by the ESPRESSO DRS pipeline. The CCFs of the reduced spectra (corrected by the barycentric Earth RV, BERV, of each observation) are computed by cross-correlating them with a weighted binary mask. We used the default ESPRESSO mask of spectral type K2, the closest to the spectral type of WASP-156. The star's systemic velocity is about 9.3~\kms \citep{gaia2018dr2}, therefore, the RV grid of the CCF was defined to be centred at that value and computed over an RV span of $\pm60$~\kms from the centre in steps of 0.5~\kms (which is the average ESPRESSO pixel size in velocity units). The cross-correlation is performed on an order-by-order basis, resulting in a disc-integrated CCF per order. A final CCF per observation is obtained by co-adding the CCFs of all orders. 
Finally, a Gaussian function is fitted to the co-added CCF. The final disc-integrated RVs are the centroids of the best-fit Gaussians to the co-added CCFs.

MAROON-X covers a redder wavelength range than ESPRESSO (up to 9200~\AA{}, while ESPRESSO goes up to 7887~\AA{}), hence, using one of the default ESPRESSO masks (which only go up to 7887~\AA{}) results in the loss of information from several red orders. To include information from the whole spectral range, we needed a different CCF mask to include all the RV information of the reddest orders. We used the \texttt{raccoon}\footnote{\url{https://github.com/mlafarga/raccoon}} pipeline \citep{lafarga2020carmenesccf} to build a mask using a high-S/N, observation-co-added template built from the MAROON-X observations themselves with \texttt{serval} \citep[see next paragraph for more information on \texttt{serval},][]{zechmeister2018serval}. We selected lines with FWHM between 2 and 20~\kms, minimum contrast of 6\%, and relative depth below 80\%. We removed lines in strongly contaminated telluric regions using a custom telluric mask from the MAROON-X site.
We then used this custom mask to compute the CCFs of all orders using a custom Python CCF code for MAROON-X.

In addition to the standard CCF-based RVs, we also computed disc-integrated RVs using the template matching approach implemented in \texttt{serval} \citep{zechmeister2018serval}. \texttt{serval} performs an iterative process consisting of first computing an initial set of RVs by performing a least-squares fit between all the observations and the observation with the highest S/N, then building a high-S/N template by co-adding the Doppler-shifted observations, and finally improving on the Doppler shift solution by recomputing the RVs with the high-S/N co-added template. We applied \texttt{serval} to both the ESPRESSO and MAROON-X reduced observations.

We note that we had to perform a relatively strong masking of telluric lines when computing the \texttt{serval} RVs for both instruments, as well as when creating the mask and CCFs for the MAROON-X data. Not removing enough telluric-affected regions of the spectrum resulted in an extra slope in the computed RVs, as reported by, e.g., \citet{Silva2025}.
\citet{Silva2025} found extra slopes in template matching RVs compared to CCF-derived RVs, especially for time series observations taken over a short time span, as is our case. The authors argue that the cause of these extra slopes in template-matching-based RVs stems from contamination from micro-tellurics and other sources of correlated noise, which could explain the slopes we initially obtained here with \texttt{serval}.
The MAROON-X CCF RVs, which initially also showed a similar slope, are computed using a mask that has been derived from a template built with \texttt{serval}. Therefore, we can expect that these RVs will also suffer from similar effects.
In our case, an aggressive removal of the telluric lines removed the slope, however, this was at the expense of losing some of the contaminated stellar lines. Other possible avenues to solve this bias, currently not available to us, would be obtaining further observations at different BERV, so that the impact of tellurics is spread over different spectral regions, or using another template of a similar star built from enough observations spread over time.

\subsection{Classical Rossiter-McLaughlin}

The classical Rossiter-McLaughlin method consists of fitting the time series of the disc-integrated RVs. During the transit occultation, these `disc-integrated' RVs only account for the unocculted part of the stellar disc, which result in the RM anomaly.

\subsubsection{\texttt{ironman}}

For this analysis, we only worked with the ESPRESSO \texttt{serval} RVs. For the current dataset, they are more precise than the CCF-derived RVs, likely because the template matching of the full spectra ends up extracting more RV information than the pre-selected lines of the CCF mask. Moreover, the classical RM effect is not detected in the MAROON-X data. Given the small RM amplitude that we measure from the ESPRESSO data and the scatter of the MAROON-X data, we do not expect to be able to measure the RM effect in the MAROON-X data alone (see Figures~\ref{fig:crm_fit_mx} and \ref{fig:crm_corner_es}).

To model the in-transit datasets, we used the \texttt{ironman}\footnote{\url{https://github.com/jiespinozar/ironman}} code \citep{Espinoza-Retamal2024}, which relies on different publicly available codes to derive stellar obliquities. Namely, \texttt{rmfit} \citep{Stefansson22} to model the in-transit RVs, \texttt{radvel} \citep{radvel} to model the Keplerian signal, and the \texttt{dynesty} dynamic nested sampler \citep{dynesty} to get the posteriors.

We placed informative Gaussian priors for almost all the parameters of the planet and its orbit, based on the values shown in Table~\ref{tab:juliet_params}. The exceptions were the sky-projected obliquity $\lambda$ and the projected rotational velocity $v\sin{i_\star}$, where we placed uninformative priors between -180 and 180$^{\circ}$, and 0 and 2 km s$^{-1}$, respectively. We also placed an informative prior for $\beta$, the intrinsic linewidth accounting for instrumental and macroturbulence broadening \citep[see][]{Hirano10}. We considered an instrumental broadening of 2.15~\kms because of the ESPRESSO resolution and a macroturbulence broadening of 2.8~\kms derived from the macroturbulence law of \citet{Valenti2005} using the effective temperature reported in Table~\ref{tab:paws}. We added the instrumental and macroturbulence broadening in quadrature to set our prior, with an uncertainty of 2~\kms. We performed the analysis for each ESPRESSO dataset independently and for both datasets together. For the joint analysis, we considered both datasets as taken with the same instrument, but allowed a different jitter term to take into account the differences in the observing conditions each night. Shared parameters between datasets are the limb darkening coefficients and the RV offset. We tested fitting different RV offsets, but found a negligible difference between the two nights. 

\subsubsection{Results}

Our fit for the first night resulted in $\lambda=-28^{+17}_{-15}\,^{\circ}$, while for the the second night we obtained $\lambda=6^{+41}_{-49}\,^{\circ}$, with the relatively large uncertainty mostly coming from the lack of transit egress observations. As the projected obliquity was consistent between datasets, we decided to formally adopt the results of the joint fit, which gave $\lambda=-15^{+17}_{-16}\,^{\circ}$. 

We derive a $\vsini=0.44\pm0.08$~\kms from the joint fit of the two nights (and $\vsini=0.52^{+0.12}_{-0.11}$ and $0.25^{+0.17}_{-0.16}$~\kms for the first and second night separately, which agree within 1~$\sigma$). This is significantly smaller than the reported SOPHIE-based value of $\sim4$~\kms (note again that the SOPHIE value is based on a calibration), but consistent with the upper limit of $2$~\kms we obtained from our \texttt{PAWS} analysis of the ESPRESSO data.

\begin{figure}
    \centering
    \includegraphics[width=\columnwidth]{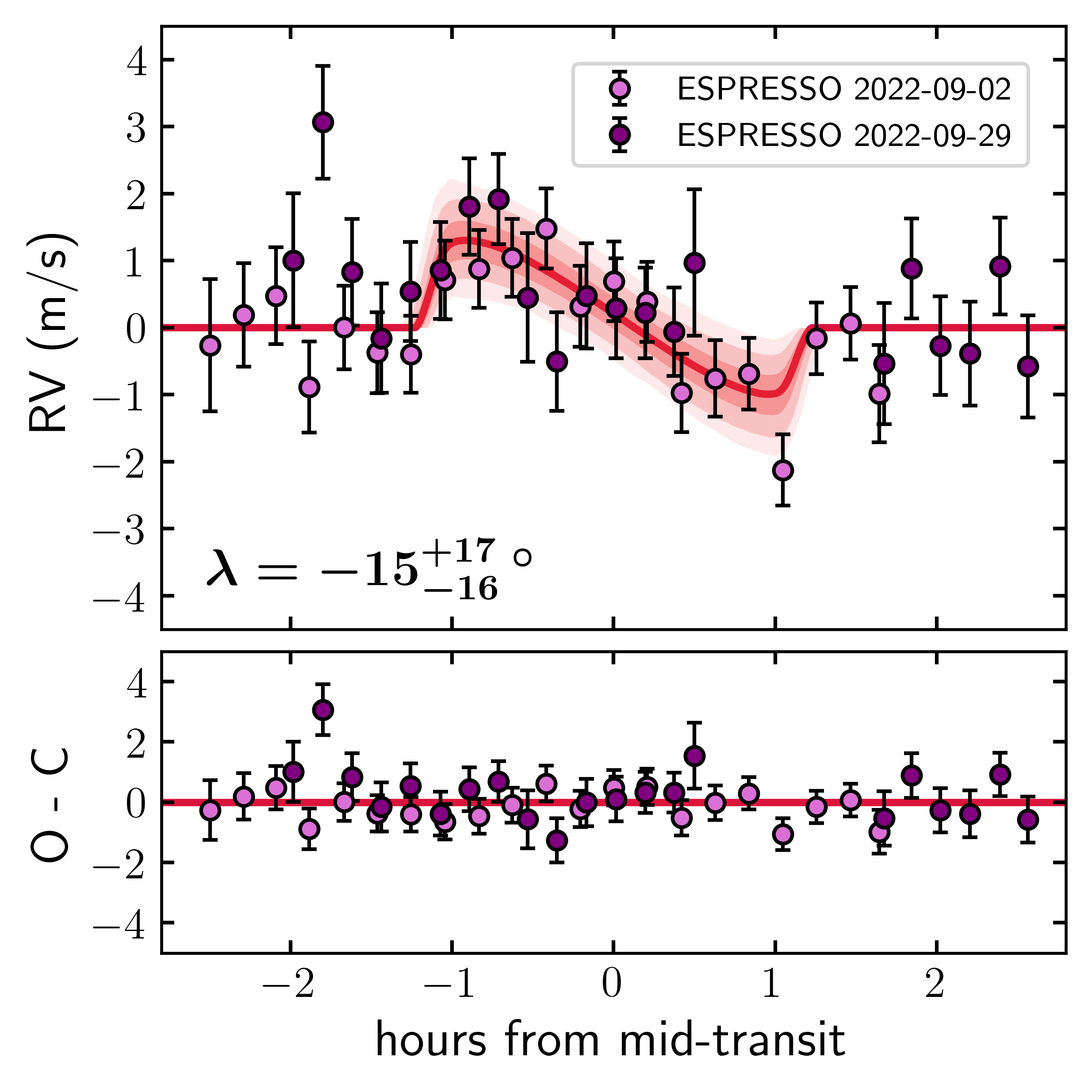}
    \caption{Disc-integrated radial velocities taken with ESPRESSO during the transit of WASP-156 b, after subtracting the Keplerian motion induced by the planet. We show in light pink and dark purple the measurements taken on two different nights. The best-fitting model is shown as a red line, with $1\sigma$, $2\sigma$, and $3\sigma$ models as the shaded regions. The best fit reveals a sky-projected obliquity $\lambda = -15^{+17}_{-16}\,^{\circ}$, suggesting that the planetary orbit is well aligned with the stellar spin.}
    \label{fig:crmfit}
\end{figure}

The posteriors for the remaining parameters of our joint fit are presented in Table~\ref{tab:ironman_params}; the data and best-fitting model are illustrated in Figure~\ref{fig:crmfit}. We show corner plots with the posterior distributions of $\lambda$ and \vsini for the different fits to data in Figure~\ref{fig:crm_corner_es}. This analysis of the classical RM effect suggests that the planetary orbit is well aligned with the equator of the host star. 

\begin{table}
    \centering
    \caption{Summary of posteriors of the \texttt{ironman} fit to the {\it classical} Rossiter-McLaughlin effect for the ESPRESSO transits.}
    \begin{tabular}{lr}
    \hline
       Parameter & Posterior \\ 
    \hline
       $\lambda$ ($^{\circ}$) & $-15^{+17}_{-16}$\\
       $v\sin{i_\star}$ (km s$^{-1}$) & $0.44\pm0.08$\\
       $P$ (d)  & $3.8361627\pm0.0000006$\\
       $t_0$ (BJD - 2,450,000) & $9058.6112\pm0.0001$\\
       $R_\mathrm{p}/R_{\star}$ & $0.0680\pm0.0007$\\
       $b$ & $0.40\pm0.05$\\
       $\rho_\star$ (kg m$^{-3}$) & $2130\pm140$\\
       $e$ & 0 (fixed)\\
       $K$ (m s$^{-1}$) & $19.4\pm0.9$\\
       \hline
       $\beta$ (km s$^{-1}$) & $3.6\pm1.9$\\
       $q_1^{\rm ESPRESSO}$ & $0.35^{+0.37}_{-0.25}$\\
       $q_2^{\rm ESPRESSO}$ & $0.41^{+0.37}_{-0.29}$\\
       $\gamma_{\rm ESPRESSO}$ (m s$^{-1}$) & $-0.15\pm0.16$\\
       $\sigma_{\rm ESPRESSO1}$ (m s$^{-1}$) & $0.02^{+0.12}_{-0.02}$\\
       $\sigma_{\rm ESPRESSO2}$ (m s$^{-1}$) & $0.04^{+0.34}_{-0.04}$\\
    \hline
    \end{tabular}
    \label{tab:ironman_params}
\end{table}

\subsection{Reloaded Rossiter-McLaughlin}\label{sec:rrm}

We also applied the {\it reloaded} Rossiter McLaughlin technique \citep{cegla2016rrm} to the new spectroscopic transit data. Rather than modelling the time series of apparent RVs, this method isolates the stellar light behind the planetary transit chord, thus mapping the path of the planet on the surface of the star.
We modelled the disc-integrated co-added CCFs, which ensure a higher S/N than directly using the observed spectra.
We treated each transit (two from ESPRESSO and one from MAROON-X) independently; therefore, we repeated this process for each of the three nights. In this section we provide a brief overview of the steps followed, and refer the reader to \citet{cegla2016rrm} for more details.

\subsubsection{Local CCFs}

\begin{figure*}
\centering
\includegraphics[width=\textwidth]{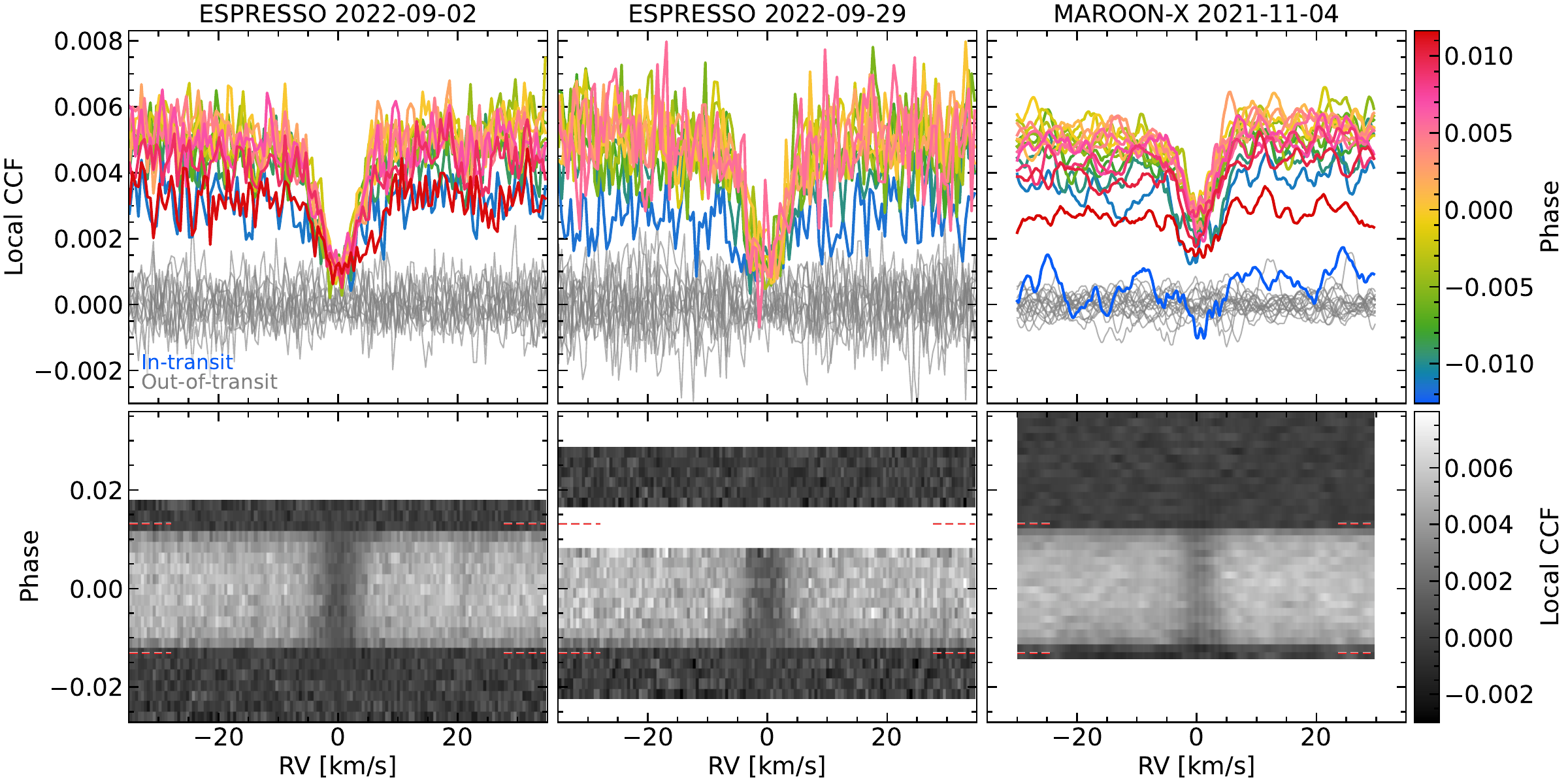}
\caption{Local CCFs, that is, profiles after subtracting each of the disc-integrated CCF from their corresponding master-out CCF (one per transit). The RV is in stellar rest frame. Left to right correspond to the first and second ESPRESSO transits, and to the MAROON-X one. The top panels show the CCF profiles (i.e. CCF flux as function of RV). The in-transit CCFs are colour-coded as a function of the orbital phase of the planet (with start-of-transit phases in blue and end-of-transit pahses in red), and out-of-transit CCFs are in grey. The bottom panel shows the same data but in a map format, with each of the CCF points colour-coded by their flux and shown as a function of the orbital phase in the y-axis. The red dashed lines indicate the start ($T_1$) and end ($T_4$) of the transits.}
\label{fig:rrmccf}
\end{figure*}

We started by Doppler-shifting the CCFs to remove the Keplerian reflex motion of the star caused by the planet using orbital parameters from Table~\ref{tab:juliet_params}.
Correcting the CCFs by the planetary RVs results into CCFs sampled at slightly different RV points. To co-add and subtract the CCFs in the next steps, we interpolated them into a common RV grid using a cubic spline.

We then co-added all the CCFs of out-of-transit observations into a master-out CCF (as mentioned above, we work with the three sets of observations independently, so we obtained one master-out CCF per transit).
In the case of the MAROON-X data, we discarded the first four out-of-transit observations, which were taken at relatively high airmass ($>2.3$) and have low S/N.
The ESPRESSO observations were all taken at airmasses below 2.3 and hence we use all of them. Preliminary tests using only pre-transit or post-transit observations to build the master-out ESPRESSO CCF did not result in a clear improvement of our results.
We measure the systemic velocity of the star by fitting a Gaussian to the master-out CCF. For the first and second ESPRESSO nights, we obtained a systemic velocity of $9.670\pm0.036$ and $9.6689\pm0.036$~\kms, consistent with literature values \citep[$9.30\pm0.61$~\kms from \textit{Gaia} DR2,][]{gaia2018dr2}. Note that this assumes that the CCF mask used by the DRS has been calibrated to be at 0 velocity. The \texttt{raccoon} mask used for the MAROON-X data has not been set to 0~\kms, hence, we do not derive a systemic velocity from that dataset.

We normalised the master-out CCF and the CCFs of individual observations by dividing them by the mean of their continuum (i.e., the region of `flat' flux outside the CCF minimum, which we define as the region outside 3.5 times the width of the CCF on either side of the CCF minimum, to make sure we are not including part of the CCF profile).
To account for the decreasing amount of light coming from the star as the planet occults it during the transit, we further normalised the CCFs by the transit light curve. We used a transit light curve model generated with \texttt{batman} \citep{batman} and parameters from Table~\ref{tab:juliet_params}.

After normalisation, we subtracted the CCF of each individual observation from the corresponding master-out, resulting in a residual CCF (see Figure~\ref{fig:rrmccf}). For the in-transit observations, the residual CCFs only contain light coming from the stellar disc region occulted by the planet. For the out-of-transit observations, the residual CCFs only contain noise. Note that we are assuming that the stellar disc does not change significantly during the time span of each transit. We refer to these residual CCFs as local CCFs because they contain stellar light from a localised area on the stellar disc.
We propagated flux uncertainties from the original disc-integrated CCFs to the local CCFs.
Finally, we fit the in-transit local CCFs with a Gaussian function to derive the local RVs, that is, the RV of the stellar disc regions occulted by the planet during its transit.

\subsubsection{Stellar surface model}\label{sec:rrm_models}

Next, we modelled the Doppler shifts of the local RVs following the models presented in \citet{cegla2016rrm}, which account for stellar rotation.
The stellar rotation along the transit chord is modelled by computing the brightness-weighted average stellar rotational velocity of the stellar disc area occulted by the planet at each observation, which depends on the projected obliquity $\lambda$ and the projected rotational equatorial velocity of the star \vsini. The model we used assumes that the star is a solid body (SB, i.e., rigid rotation).

We fitted our models with the affine-invariant ensemble sampler for Markov chain Monte Carlo (MCMC) implemented in the package \texttt{emcee} \citep{foremanmackey2013emcee}. We used uniform priors on $\lambda$ ($\mathcal{U}(-180,\,-180\degr)$) and \vsini ($\mathcal{U}(0,\,100\,\kms)$).
We used 40 walkers, 4000 total iterations, and discarded a burn-in of 300 iterations. We initialised our walkers around a multivariate Gaussian distribution with small ($10^{-7}$) width centred on initial values of $\lambda=0\degr$ and $\vsini=2$~\kms. 
Setting the starting points at different values did not affect final posterior distributions obtained.
We modelled each of the three transits separately, the two ESPRESSO transits together, and all three transits together.

Our main dataset considered all in-transit observations. Additionally, we also fitted the models to a reduced dataset where we discarded data points close to the stellar limb, with $\langle\mu\rangle<0.3$.\footnote{ where $\langle\mu\rangle$ is the brightness-weighted stellar disc position, with 0 being at the limb and 1 at disc centre, and $\mu=\cos\theta$, where $\theta$ is the centre-to-limb angle} This cut in $\langle\mu\rangle$ discards the first in-transit observation of the second ESPRESSO night, and the first and last MAROON-X in-transit observations. 
Our motivation to consider this slightly reduced dataset is that at the limb, the local CCFs have significantly less flux than closer to the disc centre (see first and last in-transit local CCFs in Figure~\ref{fig:rrmccf}), and hence the derived local RVs are less precise \citep[see e.g.][]{cegla2016rrm}.
All of the derived parameters of this reduced dataset agree well within $1\,\sigma$ with those derived from the dataset that considered all observations. Therefore, in the following, we only report our results for the dataset that considered all observations.

Note that we also considered models which, in addition to SB rotation, also include centre-to-limb convective variations. However, we found that those models overfit the data and do not necessarily improve the fit, despite being more complex. We provide details about these models in Appendix \ref{sec:rrm_extra} but do not consider them further here.

\begin{figure*}
\centering
\includegraphics[width=\textwidth]{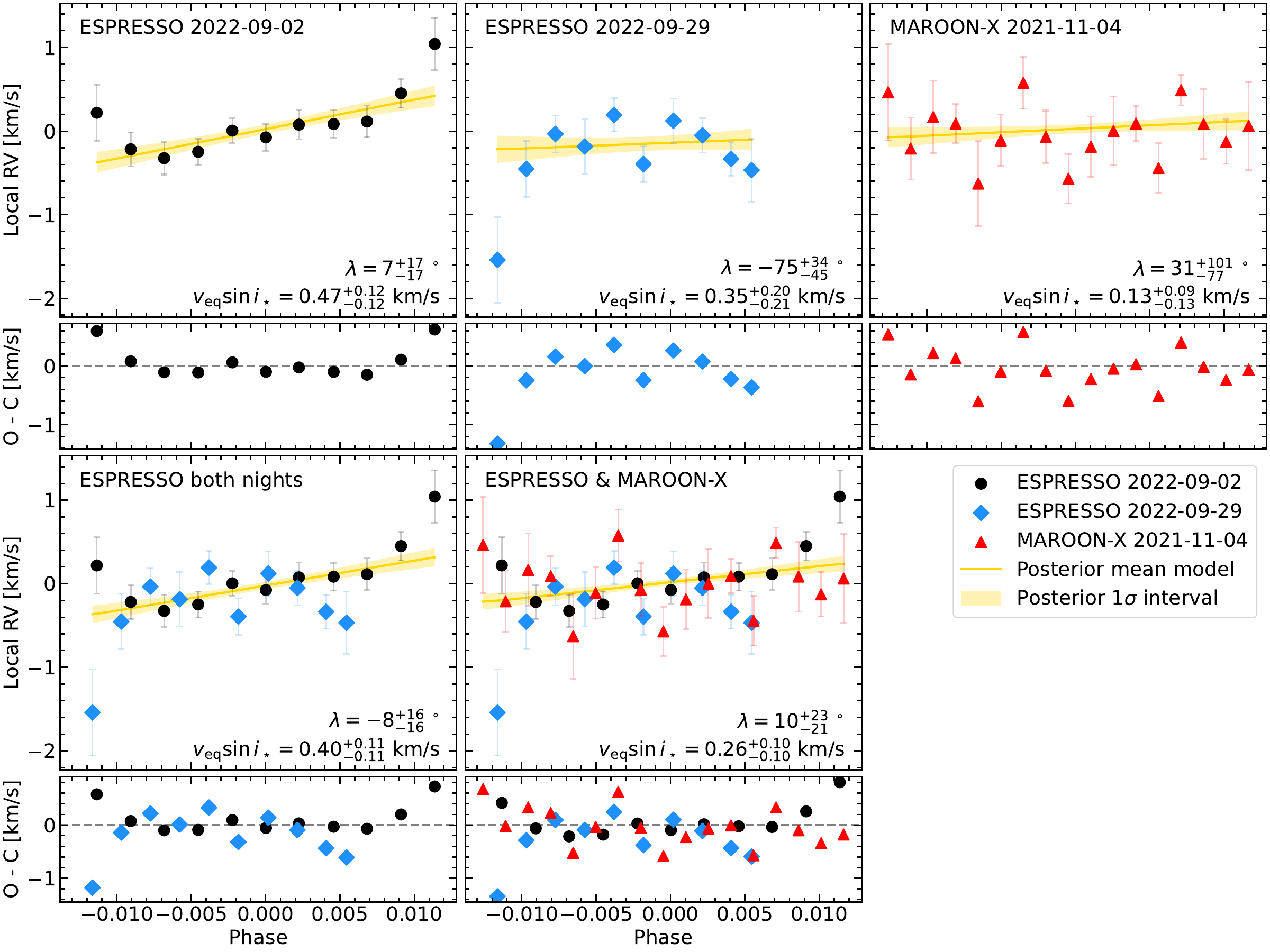}
\caption{Local RVs and preferred solid body models for the ESPRESSO (first night in black circles, second night in blue diamonds) and MAROON-X (red triangles) transits. Top panels show the fits to individual transits, and bottom panels show the fit to the two ESPRESSO transits together (left) and to all ESPRESSO and MAROON-X transits together (right). The solid yellow line represents the posterior mean model of the fit to each set of data, and the yellow shaded area the $1\,\sigma$ credibility interval around the model. The text in each panel shows the mean posterior parameters and their $1\,\sigma$ uncertainties.}
\label{fig:rrmrvfit}
\end{figure*} 

\begin{table}
\centering
\caption{Summary statistics of posterior distributions of the SB model fit of the \textit{reloaded} RM analysis for the ESPRESSO observations.}
\begin{tabular}{lcccc}
\hline
Parameter & Night 1 & Night 2 & Both nights \\
\hline
$\lambda$~(\degr) & $7^{+17}_{-17}$ & $-75^{+34}_{-45}$ & $-8^{+16}_{-16}$ \\
\vsini~(\kms) & $0.47^{+0.12}_{-0.12}$ & $0.35^{+0.20}_{-0.21}$ & $0.40^{+0.11}_{-0.11}$ \\
\hline
\end{tabular}
\begin{flushleft}
\end{flushleft}
\label{tab:results_rrm}
\end{table}

\subsubsection{Results}

The best-fitting SB models for the different data sets considered are shown in Figure~\ref{fig:rrmrvfit}, Table~\ref{tab:results_rrm} shows a summary of the posteriors for the ESPRESSO data, and we show the corner plots and posterior distributions in Figure~\ref{fig:rrm_corner_sb_esmxall}.

The first night of ESPRESSO observations results in an aligned projected obliquity of $\lambda=7^{+17}_{-17}\,^{\circ}$. 
The second night of ESPRESSO observations yields a projected obliquity of $\lambda=-75^{+34}_{-45}\,^{\circ}$, with larger uncertainties than the first night, similarly to what happens in the classical RM analysis. This value could indicate a slight misalignment, however, the uncertainties are large enough that an aligned orbit cannot be ruled out. 
The value derived from the second night agrees within $2\,\sigma$ with the one from the first night, but again, note the larger uncertainties of the second night. Each night agrees within $1\,\sigma$ with the individual night values derived from the classical RM.

The observing conditions of the second night were slightly worse, with a higher airmass at the start of the night, higher water vapour and ambient humidity during the transit and during the post-transit observations, and higher seeing during the post-transit observations. As a result, the S/N was also lower than that of the first night, although note that the observations of the second night were 100 s shorter than those of the first night (600~s vs 700~s), which also contributes to the S/N decrease. These factors result in local CCFs with lower S/N, and hence local RVs with larger uncertainties.
More importantly, the transit egress of the second night, which significantly helps constrain the obliquity, was not observed due to pointing restrictions. 

Despite these differences, the average local RVs show a strong overlap during most of the transit (except at very low $\mu$ angles and right before the second night observations had to be aborted, where we expect a low local RV precision), and the systemic velocities that we derive from the master out CCFs of each night agree well with each other, showing that there is not a large disagreement between the two datasets.

When modelling both nights together, we obtained a projected obliquity of $\lambda=-8^{+16}_{-16}\,^{\circ}$. The higher quality observations of the first night dominate the fit, and the measured $\lambda$ agrees within $1\,\sigma$ with the value derived from the first night alone, and within $2\,\sigma$ with the value from the second night. Comparing with the results from the classical RM, again the $\lambda$ values derived from the two nights together agree within $1\,\sigma$.

The local RVs extracted from the MAROON-X data are significantly noisier than the ESPRESSO RVs. We obtained a projected obliquity of $\lambda=31^{+101}_{-77}\,^{\circ}$ considering the MAROON-X data alone. The value has large uncertainties, making it unconstrained, as within $1\,\sigma$, the orbit is consistent with being aligned, polar, and retrograde.
Using the two ESPRESSO nights and the MAROON-X night together, we obtained $\lambda=10^{+23}_{-21}\,^{\circ}$. Again, the first ESPRESSO night dominates the fit, and the measured $\lambda$ agrees well within 1~$\sigma$ with the value derived from the first ESPRESSO night alone, both ESPRESSO nights, and the classical RM analysis.

As in the case of the classical RM analysis, we decided to adopt the value derived from the fit to both ESPRESSO nights, that is, $\lambda=-8^{+16}_{-16}\,^{\circ}$. Note again that this value is consistent within $1\,\sigma$ with the results derived for the first ESPRESSO night, all ESPRESSO and MAROON-X nights together, and with the classical RM analysis (for all ESPRESSO individual nights and both nights considered together).

Regarding the measured \vsini, we obtained values below $\simeq1$~\kms for almost all data sets and models considered, in agreement with the value measured with \texttt{PAWS} in Sect. \ref{sec:paws}.
In particular, for the ESPRESSO data we obtained $\vsini=0.47^{+0.12}_{-0.12}$~\kms, $0.35^{+0.20}_{-0.21}$~\kms, and $0.40^{+0.11}_{-0.11}$~\kms for the first, second, and both nights, respectively. All these values agree within 1~$\sigma$ with the \vsini measured in the classical RM analysis.
The MAROON-X data alone yielded a lower \vsini of $0.13^{+0.09}_{-0.13}$~\kms, with a posterior distribution peaking at $0$~\kms (see Figure~\ref{fig:rrm_corner_sb_esmxall}), which we consider a non-detection. The MAROON-X data also seems to affect the fit when considering all nights together, with $\vsini=0.26^{+0.10}_{-0.10}$~\kms. 
For the \vsini we also adopt the value derived from the fit to the two ESPRESSO nights together, $0.40^{+0.11}_{-0.11}$~\kms.


\section{Discussion}\label{sec:discussion}

\subsection{Obliquity and rotational velocity values}\label{sec:discusionvalues}

\citet{bourrier2023dream} estimated the projected stellar obliquity of the system to be $\lambda=105.7^{+14.0}_{-14.4}\,^{\circ}$ using one transit obtained with the visible arm of CARMENES.
In comparison, here, our analyses of the new ESPRESSO transits suggest that the obliquity is instead consistent with an aligned orbit ($\lambda=-15^{+17}_{-16}\, ^{\circ }$ from the classical RM and $\lambda=-8^{+16}_{-16}\, ^{\circ }$ from the reloaded RM), and is not compatible with a polar orbit.

The RM analysis of the CARMENES data presented in \citet{bourrier2023dream} clearly detects the planet-occulted CCFs (i.e. equivalent to the local CCFs).
As explained in \citet{bourrier2023dream}, a preliminary fit to the CARMENES local CCFs performed with broad uninformative priors results in a broad posterior for the \vsini, peaking at 0~\kms and $<6$~\kms at 3$\sigma$, and a posterior for $\lambda$ peaking at $\sim100\, ^{\circ }$ with shallow wings spanning the whole parameter space. 
A subsequent fit constraining the \vsini prior to a normal distribution following the value of $3.8\pm0.9$~\kms from \citet{demangeon2018wasp} removes the broad wings on the $\lambda$ posterior, yielding a well-defined value of $105.7^{+14.0}_{-14.4}\, ^{\circ }$.
In general, RM measurements can suffer from degeneracies between $\lambda$ and \vsini, especially if the impact parameter $b$ is close to 0, and/or if the amplitude of the RM signal is small compared to the RV uncertainties \citep[e.g.,][]{Albrecht2011,brown2012rm2,brown2017rm}. \planet does not transit the centre of the stellar disc ($b\sim0.4$), however the RM signal is relatively small, making a precise measurement of $\lambda$ challenging, especially with the low S/N data obtained with CARMENES.

Our re-analysis of the stellar parameters of WASP-156 using the ESPRESSO data and the \texttt{PAWS} pipeline (see Sect. \ref{sec:paws}) shows that the \vsini is likely $\leq2$~\kms. Hence, the polar obliquity derived from the CARMENES data relied on an inaccurate prior for the \vsini. 
This low \vsini value also agrees with the \vsini estimated from both the classical ($\vsini=0.44\pm0.08$~\kms) and reloaded RM ($\vsini=0.40\pm0.11$~\kms) analyses presented here using the new data. 
Note that due to the spatial information of the stellar disc that we extract from the RM effect, our analyses can measure extremely low \vsini values \citep[lower than the intrinsic line profile of the spectrograph;][]{CollierCameron2010tomography1}, but we still would advise caution with the specific \vsini values derived from RM analyses.

Although it seems like the restrictive prior on the (inaccurate) \vsini helped constrain the polar orbit in \citet{bourrier2023dream}, their initial fit with uninformative priors still resulted in a polar orbit, albeit with looser constraints. Therefore, there could be other factors in play that led to that result.
According to \citet{bourrier2023dream}, the WASP-156 CARMENES data suffers from strong telluric contamination. Out of three transits taken on different nights, two were discarded due to clear telluric contamination or spurious signals in the CCFs, and the night used in the RM analysis still showed residuals correlated with airmass, also probably due to tellurics. 
The wide posterior distribution obtained with the initial unconstrained fit to the CARMENES data means that there is a non-negligible probability that the orientation of the transit chord (constrained through $\lambda$ and $\vsini$, that is, the RM signature itself), was not constrained with those data. This would imply then that the signal at $\lambda\sim100\, ^{\circ }$, which the subsequent fit with the restrictive prior on \vsini makes more likely, could be a spurious signal from systematics in the data.
The telescopes used here and in \citet{bourrier2023dream} also have different collecting power, with CARMENES mounted on a 3.5~m telescope and ESPRESSO and MAROON-X in 8~m class telescopes. This led to observations with higher S/N and better time sampling for the new data presented here, which could also have an effect in explaining the different constraints in the derived orbit.

Recently, \citet{Jiang2026} used the ESPRESSO data obtained on 2022-09-02 to measure the obliquity of the system using the classical RM method, obtaining $\lambda=-5^{+33}_{-35}\, ^{\circ }$ and $\vsini=0.64^{+0.16}_{-0.33}$~\kms, which agree within $1\,\sigma$ with our results but are significantly less precise. Note that \citet{Jiang2026} used CCF-derived RVs from the data reduced with DRS version 2.5.0 and adopted literature system parameters \citep[from][]{demangeon2018wasp}, in particular, a smaller impact parameter $b$ than the one we derived here, which can introduce a degeneracy between \vsini and $\lambda$, increasing uncertainties. 
To understand this difference, we repeated our RM analyses of the first ESPRESSO night using the same impact parameter as in \citet{Jiang2026} and also using CCF-derived RVs. 
In particular, we applied the classical RM to (i) CCF-derived RVs using the same orbital parameters as in \citet{Jiang2026}, (ii) CCF-derived RVs using our newly derived parameters, and (iii) template-matching derived RVs with the \citet{Jiang2026} parameters. For all cases we obtained an aligned orbit with $\lambda$ uncertainties of $\pm30\degr$ for case (i), and $\pm25\degr$ for cases (ii) and (iii). Case (i) is the most similar to the analysis presented in \citet{Jiang2026} and we obtained similar uncertainties. For the other two cases, the uncertainty decrease is likely driven by the larger impact parameter of our new orbit (ii) and the better precision of the RVs derived with template-matching compared to those from the CCFs (iii). As a reminder, our original classical RM fit with the template-matching RVs and new orbital parameters yields an aligned orbit with $\pm16\degr$ uncertainties.
For completeness, the reloaded RM analysis using the \citet{Jiang2026} parameters also results in an uncertainty increase, $\pm30\degr$.
We conclude that our results improved on the uncertainties due to the new orbital parameters (increased impact factor) and the better precision of the \texttt{serval} RVs compared to the CCF ones.

Our analyses indicate that the projected obliquity of the planet is aligned rather than polar. However, the true 3D obliquity $\psi$ of the system could still be polar if the star has an inclination close to pole on. 
Currently, the stellar inclination of WASP-156 is unknown and challenging to constrain. The star has a non-detected rotation period and a relatively old age \citep[e.g. $\sim5.6$~Gyr,][]{MacDougall2023}, which could be consistent with a low equatorial rotational velocity, implying a stellar inclination close to edge on. However, the measured \vsini is so low, that even inclinations moderately close to pole on would still result in relatively low equatorial rotational velocities.
Since the stellar inclination is unknown, to estimate the possible $\psi$ of the system we simulated 10000 random system configurations by sampling an isotropic distribution of stellar inclinations $i_\star$ (uniform $\cos i_\star$ between 0 and 1), and Gaussian distributions based on our values of $\lambda$ ($-15^{+17}_{-16}\degr$ from the classical and $-8^{+16}_{-16}\degr$ from the reloaded RM) and orbital inclination ($i_\mathrm{p} = 88.1\pm0.3\degr$), and compute $\psi$ as $\cos\psi = \sin i_\star \cos\lambda \sin i_\mathrm{p} + \cos i_\star \cos i_\mathrm{p} $ \citep[e.g.,][]{Fabrycky2009}. This results in a $\psi$ distribution with median and $1\,\sigma$ uncertainties of $37^{+22}_{-18}\degr$ and $34^{+23}_{-18}\degr$ for the classical and reloaded RM $\lambda$ values, respectively (see Fig. \ref{fig:psi}). Therefore, given a randomly distributed stellar inclination, it is unlikely that the true obliquity is polar.

\subsection{Search for companions}\label{sec:comp}

\begin{figure*}
    \centering
    \includegraphics[width=\textwidth]{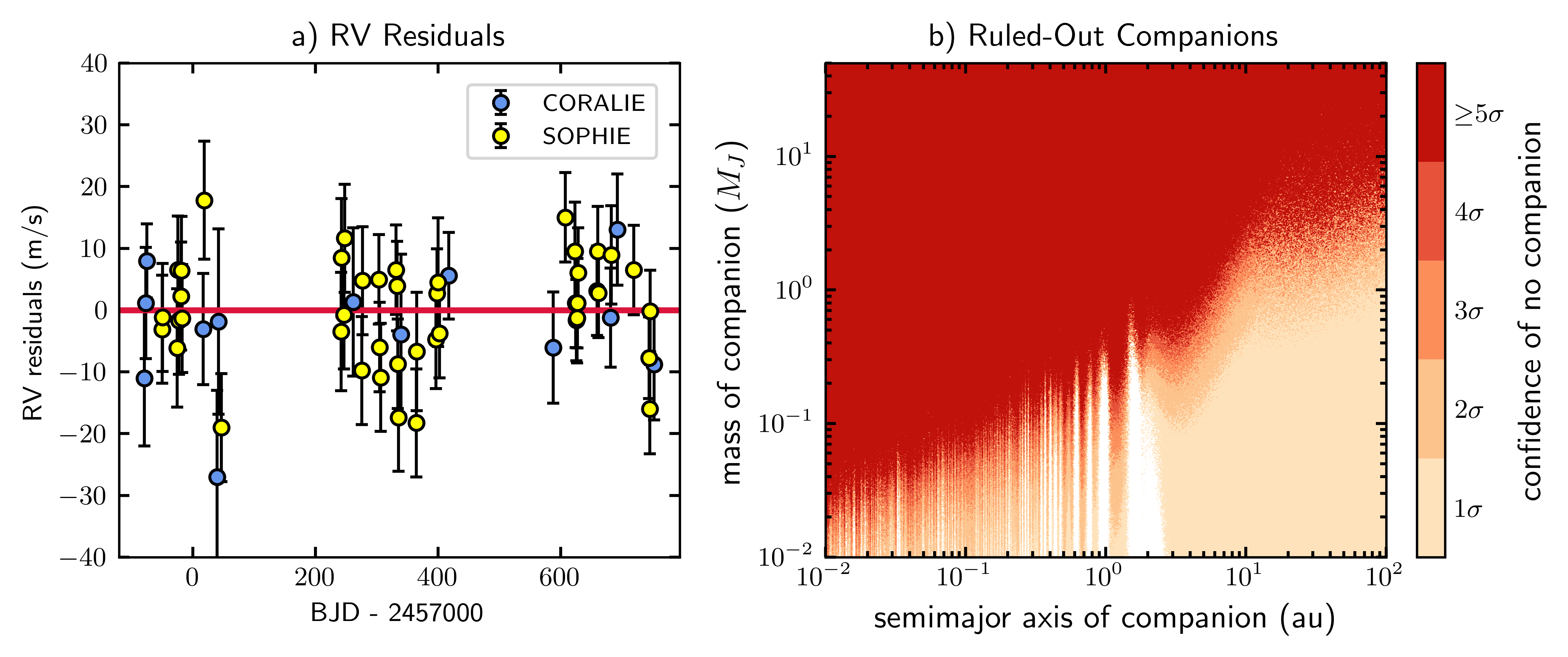}
    \caption{a) RV residuals from the \texttt{juliet} fit as a function of time. Different colours represent different instruments used to observe the star. No signals of additional companions are identified. b) Mass versus semimajor axis diagram of companions that can be ruled out thanks to the lack of additional signals in the RV residuals. Different colours represent different confidence levels. Based on the 2.4 years of RV observations, we can rule out Jupiter-mass companions to WASP-156 within 5~au at a $5\sigma$ confidence.}
\label{fig:companions}
\end{figure*}

In this section we explore the potential presence of companions in the system, which is key information to understand the system's formation and evolution.
High-resolution imaging observations of WASP-156 rule out stellar companions within $\sim25$ to 500~au, down to companions $\sim3-5$ to 7~mag fainter than WASP-156. 
In particular, WASP-156 was observed with the Zorro speckle imaging instrument on Gemini South on 2020/11/28 \citep{Scott2021}, which performs simultaneous speckle imaging in two bands (centred at 562~nm and 832~nm) and provides contrast curves out to $\sim1$~arcsec 
\citep{Howell2011}. Results for WASP-156 (as available on ExoFOP, PI: Howell, see Fig.~\ref{fig:zorro}) show that it does not have companions brighter than $\sim$5 to 7~mag below that of WASP-156 from 0.2~arcsec (the diffraction limit) out to 1.2~arcsec (corresponding to a projected physical separation of $\sim25$ and $\sim150$~au at the distance\footnote{$\sim122$~pc from \textit{Gaia} DR3} of WASP-156).
Dressing et al. (subm.) observed WASP-156 on 2019/07/21 using the ShARCS camera on the Shane 3-m telescope at Lick Observatory \citep{2012SPIE.8447E..3GK, 2014SPIE.9148E..05G, 2014SPIE.9148E..3AM}, using the Shane adaptive optics system in natural guide star mode and a $Ks$ filter ($\lambda_0 = 2.150$ $\mu$m, $\Delta \lambda = 0.320$ $\mu$m), and reduced the data using the publicly available \texttt{SImMER} pipeline \citep{2020AJ....160..287S}.\footnote{\url{https://github.com/arjunsavel/SImMER}} The observations achieved a magnitude contrast of 3.2 at 0.5~arcsec, 5.0 at 1.0~arcsec, down to below 7 at 4~arcsec (corresponding to a projected physical separation between $\sim60$ and $\sim500$~au at the distance of WASP-156), finding no nearby stellar companions within their detection limits.
On wider scales, WASP-156 is not listed in the wide binary catalogue of \citet{El-Badry2021}, which is based on Gaia Early DR3 proper motions and parallaxes \citep{Gaia_EDR3}, indicating the absence of co-moving stellar companions out to separations of $\sim$1~pc. 
The Gaia DR3 Renormalised Unit Weight Error \citep[RUWE,][]{gaia2023dr3} value of $\sim$1.1 indicates a well-behaved single-star astrometric solution, with no evidence for significant perturbations from an unresolved companion. 
Altogether, all these constraints strongly support a single-star configuration for WASP-156.

In addition to stellar companions, we also searched for additional planetary companions in the system. The residuals of the fit presented in Section~\ref{sec:refit} do not show any significant periodic signals or long-term RV trends (Figure~\ref{fig:companions}), suggesting the absence of massive close-in or intermediate-separation planets. Following the approach of \citet{Espinoza-Retamal2024}, we performed a population synthesis analysis to quantify the types of planetary companions that can be ruled out by our data. In brief, we generated a synthetic population of companions with masses between 0.01 and 50~$\Mjup$ and semimajor axes between 0.01 and 100~au, assuming isotropic inclinations and eccentricities drawn from the distribution of \citet{Kipping2013_ecc}, with random arguments of periastron. The RV signals induced by these synthetic companions were then compared to the observed RV residuals.
By fitting the unknown instrumental offsets in Sect. \ref{sec:refit}, we could have potentially removed long-term trends due to companions from the data. Hence, here we only use the SOPHIE and CORALIE data, which span the longest baseline and overlap.
Figure~\ref{fig:companions} shows the regions of the mass versus semimajor axis plane that can be excluded at different confidence levels. 
With a baseline of $\sim2.4$ years, our RV data rule out the presence of Jupiter-mass companions within $\sim$5~au at the $5\sigma$ confidence level.

\subsection{Possible formation mechanisms}

The close-in, aligned, and nearly circular orbit of WASP-156~b ($e<0.16$ at $3\sigma$) can be naturally explained if the planet formed and evolved within a protoplanetary gas disk that was aligned with the stellar equator. In this scenario, both in-situ formation \citep[e.g.,][]{Batygin2016,Boley2016} and inward migration driven by nebular tides \citep[e.g.,][]{Goldreich1980,Lin1986} provide viable pathways to produce a short-period, well-aligned Neptune.

However, protoplanetary discs aligned with the stellar spin are not necessarily expected, and several studies based on simulations and observational data show that primordial disc misalignments can be common \citep[e.g.][]{Bate2010,Lai2011,Thies2011,Fielding2015,Bate2018,Hurt2023,Biddle2025}.
On the other hand, young exoplanets have been found to be mostly aligned \citep[e.g.][]{Montet2020,Hirano2020,Martioli2020,Palle2020,Heitzmann2021,Wirth2021,Johnson2022,Franson2023,Gan2024,Hirano2024}, and most compact multi-planet systems, thought to preserve primordial alignments and migrate inwards via disc migration, have also been found in aligned configurations \citep[e.g.][]{Sanchis-Ojeda2015,Anderson2015,Wang2022,Knudstrup2024,Zak2024,Radzom2025}.
Despite that, the sample of known young exoplanets and compact systems is still small and exceptions to the commonly observed alignment exist \citep[e.g.][]{Huber2013,Dalal2019,Yu2025}.

The apparent contradiction between these two results regarding formation can be resolved if primordial misalignments between the star and disc mainly impact the outer disc regions, while the inner parts of the disc, where planets form, remain aligned \citep[e.g.][]{Marino2015,Casassus2018,Francis2020,Ansdell2016}.
If WASP-156~b formed in an inner disc region that remained aligned with the stellar spin, its current aligned and circular orbit follows naturally from disc-driven evolution.

High-eccentricity tidal migration offers, in principle, an alternative route to produce a short-period, circular orbit. In this framework, the planet is initially placed on a highly eccentric orbit by dynamical interactions \citep[e.g.,][]{Fabrycky2007,Naoz2011,Beauge2012,Petrovich2015}, and subsequently circularises through tidal dissipation. Following \citet{Goldreich1966} and \citet{Hut1981}, we estimate an orbital circularization timescale of $\sim0.1$–$10$~Gyr, assuming a modified planetary tidal quality factor \citep[e.g.,][]{Ogilvie2007} of $Q^{\prime}_p = 10^4$–$10^6$. Given the estimated system age of $\sim5.6$~Gyr \citep{MacDougall2023}, this implies that tidal circularization from a highly eccentric orbit is, in principle, feasible and depends on the exact value of the tidal quality factor.

However, the absence of detected long-period companions capable of exciting such extreme eccentricities makes this scenario dynamically unlikely. Moreover, high-eccentricity migration pathways are generally associated with the production of misaligned orbits, in contrast with the observed alignment of WASP-156~b. Following \citet{Albrecht2012}, we estimate a tidal realignment timescale of $\sim3\times10^{4}$~Gyr, far exceeding the age of the Universe. This is expected for a relatively low-mass planet that does not raise significant tides on its host star, and strongly suggests that the system did not undergo a phase of large spin-orbit misalignment. Typical high-eccentricity migration mechanisms are therefore disfavoured. The system could, however, remain consistent with coplanar high-eccentricity migration \citep{Petrovich2015}, which preserves spin-orbit alignment throughout the dynamical evolution. This mechanism will require the existence of a massive Jupiter companion on an eccentric and aligned orbit at more than $\sim5$ au to be consistent with the constraints from long-term RV observations we derived here.

Another potential indicator of high-eccentricity migration is a large difference between a star's isochronal and gyrochronological ages. For WASP-156, \citet{demangeon2018wasp} derived an isochronal age of $6.50\pm4.03$~Gyr and a much younger gyrochronological age of $0.58^{+0.51}_{-0.31}$~Gyr using the relation from \citet{Barnes2007}. As mentioned above, recent work by \citet{MacDougall2023} also obtained a relatively old isochronal age of $5.6^{+3.9}_{-3.2}$~Gyr. The gyrochronological age was derived by assuming a stellar rotation period of $\simeq12.6$~d, computed from the reported $\vsini\simeq3.8$~\kms (using the stellar radius and assuming edge-on stellar inclination). 
A gyrochronological age shorter than an isochronal age can be explained by transfer of angular momentum from the planet to the star during tidal circularisation occurring due to high-eccentricity migration. Hence, \citet{demangeon2018wasp} proposed that a gyrochronological age younger than an isochronal age is a potential signature of high-eccentricity migration. This age discrepancy would also support the existence of a polar orbit induced by high-eccentricity migration \citep{bourrier2023dream}.
Here, however, we estimated the \vsini of the star to be much lower, $\leq2$~\kms from our stellar characterisation with \texttt{PAWS}, and $\simeq0.4$~\kms from our RM analyses. These lower \vsini values result in slower rotation periods of $\gtrsim 24$ and $\simeq120$~d, which, also following \citet{Barnes2007}, correspond to a gyrochronological age of approximately 2~Gyr and an age longer than that of the Universe, respectively. These values are now more in line with the isochronal age, and there is no need for a transfer of angular momentum to increase the stellar rotation. This picture is now compatible with disc migration or in situ formation, and also aligns well with our new obliquity measurement.

\subsection{WASP-156 in context}

\subsubsection{Obliquity bi-modality?}

\begin{figure}
\centering
\includegraphics[width=0.49\textwidth]{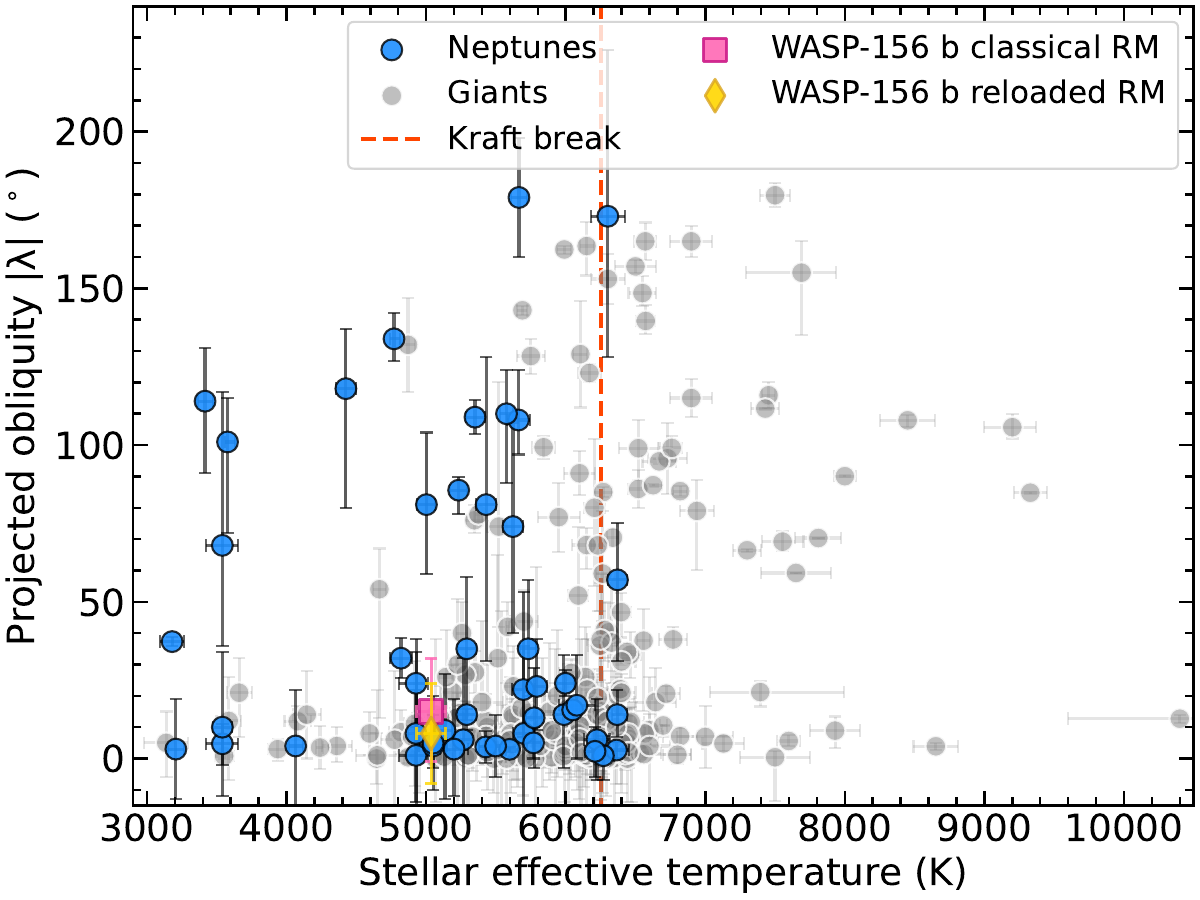}
\caption{Neptune and giant planets with known projected obliquity $\lambda$ with and uncertainty better than $50\degr$ from the TEPCat catalogue (note that we have removed the previously measured polar obliquity for \planet) as a function of stellar effective temperature. Blue circles correspond to Neptunes ($2\leq R_p \leq 8$~\Rearth and/or $10\leq M_p \leq 50$~\Mearth), grey circles show giant planets ($R_p > 8$~\Rearth and/or $M_p > 50$~\Mearth), and the pink square and yellow diamond show the new obliquity measured for \planet, from the classical and reloaded RM analyses, respectively. The red vertical dashed line shows the Kraft break at $\sim6250$~K \citep{Kraft1967}.}
\label{fig:teff_lambda}
\end{figure}

\citet{albrecht2021preponderance} suggested the existence of a bimodality in the 3D obliquity distribution ($\psi$) in a sample of 57 systems, including some Neptunes, with a preference for aligned and polar orbits (see, e.g., Figure~\ref{fig:teff_lambda}, although note that we show the distribution of projected obliquities $\lambda$ rather than the 3D obliquity $\psi$). However, there is still a debate on whether this dichotomy is real or not. For example, \citet{bourrier2023dream} also found a large fraction of polar planets, while \citet{dong2023preponderance}, \citet{siegel2023preponderance}, and \citet{Rossi2026} did not find strong evidence for a general clustering at $\sim90^{\circ}$. 
High obliquities, especially for giant planets, have also been linked to stellar hosts with effective temperatures above the Kraft break \citep[$\Teff\sim6250$~K,][]{Kraft1967}, see Fig.~\ref{fig:teff_lambda}, with thin or negligible outer convective zones unable to realign misaligned planets \citep[e.g.][]{Winn2010,Albrecht2012,Triaud2018,Mancini2022,Rossi2026}. Note that the exact value of the Kraft break has a dependency on metallicity \citep[e.g.,][]{AmardMatt2020,Amard2020,Spalding2022}.

Recently, \citet{Espinoza-Retamal2024}, \citet{Knudstrup2024}, and \citet{Rossi2026} showed that the preference for polar orbits is particularly strong among Neptunes, although, given the limited number of measurements, these results are only tentative. Note that \citet{Espinoza-Retamal2024} and \citet{Knudstrup2024} included WASP-156~b among the polar Neptunes. When considering a larger sample of Neptunes and using our updated stellar obliquity measurement for WASP-156, \citet{Espinoza-Retamal2026} found no significant evidence for a preference for polar orbits, although still with large uncertainties due to the small sample size. More stellar obliquity measurements would be necessary to continue testing the preference for polar orbits. Also, compared to giant planets, there is not a large number of Neptunes with a measured obliquity around hot ($\Teff\gtrsim6250$~K) stars (see e.g. Fig.~\ref{fig:teff_lambda}). A larger sample of Neptunes around hot stars can be proven with studies that estimate the obliquity distribution of transiting planets from \vsini measurements. Such studies suggest that small planets around hot stars tend to have higher obliquities than small planets around cooler stars, similar to the trend initially observed for giant planets \citep{Mazeh2015,Winn2017,Louden2021,Louden2024}.
However, \citet{Louden2024} show a potential dependency of this result on orbital period, with planets with $P<10$~d being in general more aligned, and planets with $P>10$~d showing more differences between planets around hot and cool stars, resulting in a wider obliquity distribution.

Our new obliquity measurement of \planet moves the planet from the tentative cluster of polar obliquities to the more abundant group of planets with aligned orbits (see Figure~\ref{fig:teff_lambda}). The host WASP-156 is a relatively cool star ($\Teff=5036\pm104$), and the aligned obliquity we measure here for \planet seems to follow the tentative trend that planets tend to be more aligned for cool stars, potentially for orbital periods $<10$~d \citep{Louden2024}. However, the sample of Neptunes with measured obliquities in this regime of the parameter space ($P<10$~d and $\Teff<6250$~k) only reaches 24 to date (16 with obliquity uncertainties better than 20\degr), which is not large enough to draw population-level conclusions, and more results are needed to confirm any trends.

\subsubsection{Short-period Neptunian landscape}

\begin{figure}
\centering
\includegraphics[width=0.49\textwidth]{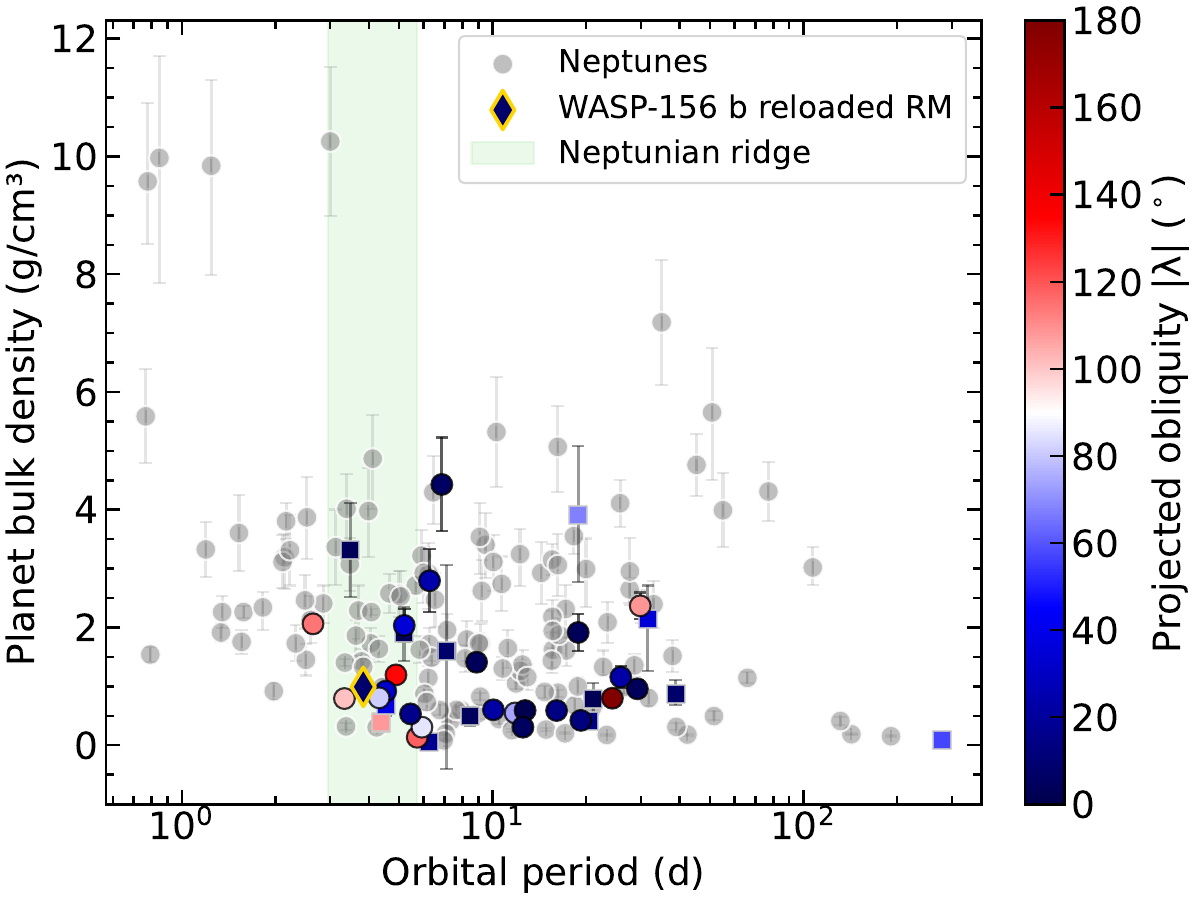}
\caption{Planet bulk density as a function of orbital period for Neptunes ($2\leq R_\mathrm{p} \leq 8$~\Rearth and/or $10\leq M_\mathrm{p} \leq 50$~\Mearth) from the TEPCat catalogue. 
Grey circles show planets without known projected obliquity and relative uncertainty in mass, radius, and density $\leq20$\%.
Planets with known projected obliquity $\lambda$ with uncertainty better than $50\degr$ are colour-coded as a function of $\lambda$, with circles representing planets with relative uncertainty in mass, radius, and density $\leq20$\%, and squares, the rest of planets in the catalogue.
\planet is represented as a diamond with yellow outline colour-coded with the reloaded RM $\lambda$ value we measure here.
We also show the approximate region of the Neptunian ridge as a shaded green region \citep[from 2.95 to 5.7~d, following ][noting that the ridge is a feature specific to the Neptunian population with radii $R_\mathrm{p}\sim4.5-8.0~\Rearth$, while this plot additionally shows smaller Neptunes]{Castro-Gonzalez2024}. 
}
\label{fig:porb_dens}
\end{figure}

\begin{figure}
\centering
\includegraphics[width=\columnwidth]{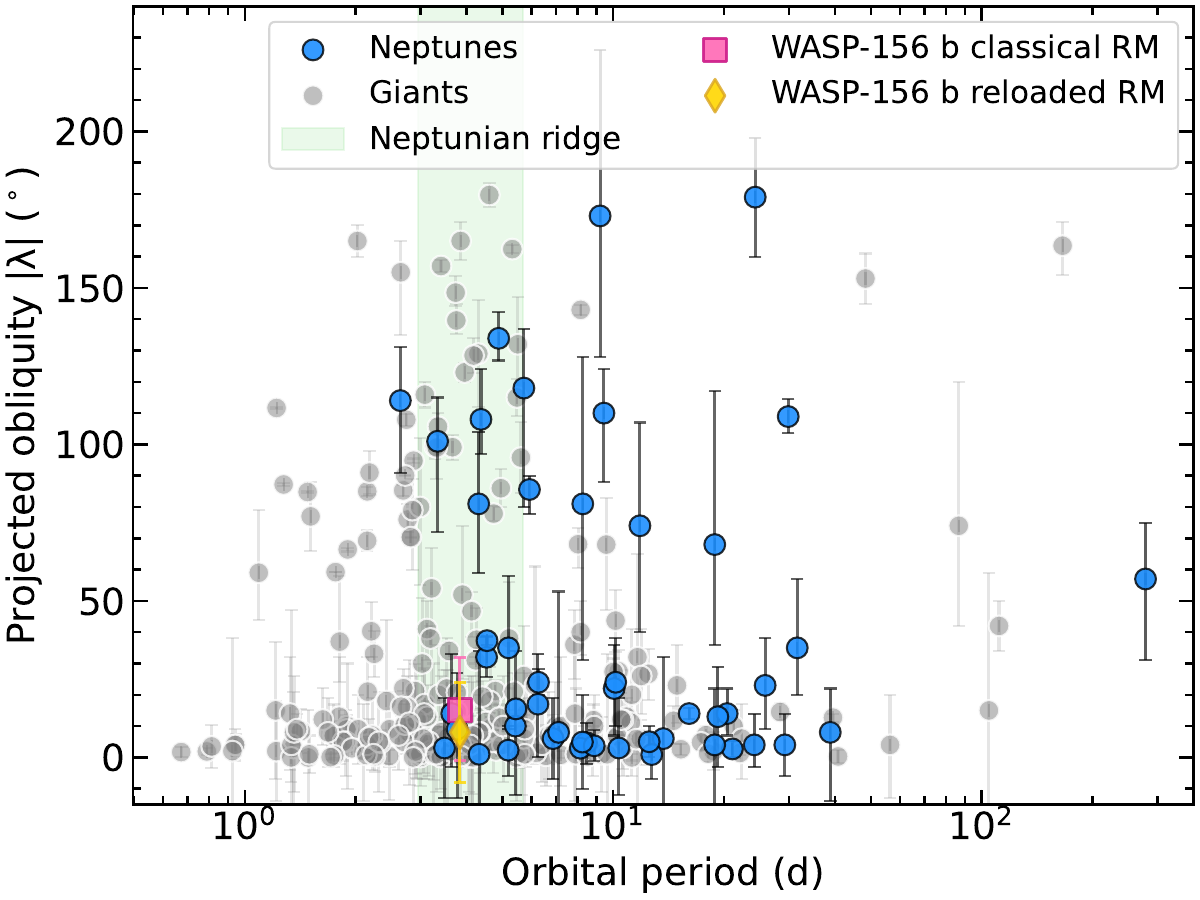}
\caption{Neptune and giant planets with known projected obliquity $\lambda$ with and uncertainty better than $50\degr$ from the TEPCat catalogue (note that we have removed the previously measured polar obliquity for \planet) as a function of orbital period. Blue circles correspond to Neptunes ($2\leq R_\mathrm{p} \leq 8$~\Rearth and/or $10\leq M_\mathrm{p} \leq 50$~\Mearth), grey circles show giant planets ($R_\mathrm{p} > 8$~\Rearth and/or $M_\mathrm{p} > 50$~\Mearth), and the pink square and yellow diamond show the new obliquity measured for \planet, from the classical and reloaded RM analyses, respectively. We also show the approximate region of the Neptunian ridge as a shaded green region \citep[from 2.95 to 5.7~d, following ][noting that the ridge is a feature specific to the Neptunian population with radii $R_\mathrm{p}\sim4.5-8.0~\Rearth$, while this plot contains smaller Neptunes and giant planets]{Castro-Gonzalez2024}.
}
\label{fig:porb_lambda}
\end{figure}

Recently, \citet{castro-gonzalez2024toi5005} highlighted that the Neptunian savanna tends to have low-density planets ($\rho\sim0.5$~g cm$^{-3}$) while ridge planets tend to be high-density ($\rho\sim1.5-2.0$~g cm$^{-3}$), suggesting different evolutionary processes for the planets in the two regimes, divided by a density of $\rho\sim1.0$~g cm$^{-3}$ note that this trend holds for planets with radii $R_\mathrm{p}\sim4.5-8.0~\Rearth$, but not for smaller Neptunes).
Expanding on this idea, \citet{bourrier2025atreides} proposed a unified view to explain the close-in Neptunian landscape integrating migration mechanisms, photoevaporation, and planetary densities. The authors propose that disc-driven migration primarily affects low-density Neptunes, which migrate inwards early-on, preserving their primordial obliquities and circular orbits. Those low-density Neptunes in the desert and ridge would be fully eroded due to the high irradiation from their stellar hosts, and only those in the savanna, further away from the star, would survive. This should result in an observable density `brink' along the ridge, with densities decreasing with increasing period (see e.g. the lower envelope in Figure~\ref{fig:porb_dens}). On the other hand, high eccentricity migration, acting during a wide range of ages, would mainly affect high-density planets, bringing them preferentially to the desert or ridge, potentially in eccentric and misaligned orbits. Hence, only high-density planets and/or late migrators would survive within the desert.

The sample of Neptunes with obliquities is still small but, as can be seen in Figure~\ref{fig:porb_lambda}, most Neptunes in the savanna tend to have aligned orbits (with few exceptions), while planets in the ridge seem to show a wider diversity of obliquities, including several polar cases. There is not a large number of desert planets with measured obliquities.

\planet is a ridge planet that has a density of $\rho\simeq1$~g cm$^{-3}$, which places it at the boundary between low- and high-density planets in the picture proposed by \citet{bourrier2025atreides}. It sits close to the proposed density brink (Figure~\ref{fig:porb_dens}), suggesting that it is sufficiently dense to have survived photoevaporation if it migrated early-on via disc-driven migration. This picture is also consistent with the low eccentricity and obliquity we measure here.

Photoevaporation is thought to play a key role in shaping the close-in Neptunian planet population \citep[e.g.][]{davis2009evaporation,Owen2013,owen2018desert}. X-ray and UV radiation from the host star can result in an expansion of the planetary atmospheres and a subsequent atmospheric escape, with the effect being strongest during the star's early, more active phase. The number of atmospheric escape studies on Neptunes is still limited; hence, it is still unclear when photoevaporation occurs and if it affects Neptunes with specific orbital distances, densities, and/or obliquities.
So far, only Neptunes within or very close to the ridge have shown clear signs of photoevaporation (escaping neutral H detected in the Lyman-$\alpha$ line or neutral He detected in the metastable 1083~nm triplet),\footnote{\url{https://research.iac.es/proyecto/exoatmospheres/table.php}} and all are in polar orbits: GJ~436~b \citep{Ehrenreich2015,Bourrier2022}, GJ~3470~b \citep{bourrier2018GJ3470,Stefansson22}, and HAT-P-11~b \citep{Allart2018,Winn2010hatp11,Sanchis-Ojeda2011hatp11,bourrier2023dream}. For the Neptune HAT-P-26~b, also within the ridge and in a polar orbit \citep{luo2026hatp26}, current data suggest contradictory results regarding its atmospheric mass-loss \citep{Vissapragada2022,Orell-Miquel2025}, and TOI-942~c, an aligned Neptune in the savanna, shows tentative signs of photoevaporation \citep{Teng2024}.

For \planet, \citet{Jiang2023} report a lack of atmospheric features from one transit observed in the optical, long-slit spectroscopic mode of the Optical System for Imaging and low Resolution Integrated Spectroscopy on the 10~m Gran Telescopio Canarias \citep[GTC OSIRIS, $R\sim1000$;][]{Cepa2000}. The observations are reported to be affected by strong instrument systematics, which reduce the S/N of the transmission spectra, leading to a non-detection. 
\citet{Orell-Miquel2025} do not detect escaping He in two transits observed in the near-infrared with the Habitable-zone Planet Finder \citep[HPF, $R=55\,000$;][]{Mahadevan2012,Mahadevan2014} on the 10~m Hobby–Eberly Telescope \citep{Ramsey1998,Shetrone2007}. They place a $3\,\sigma$ upper limit to the expected absorption of 0.9\%.
\citet{Jiang2026} looked for absorption signals of several species on the ESPRESSO data obtained on 2022-09-02 using both the transmission spectrum directly and the cross-correlation function. The authors also find a featureless transmission spectrum.
Future observations of the atmosphere of \planet could help determine whether or not the planet is affected by photoevaporation.


\section{Conclusion}\label{sec:conclusion}

We analysed new spectroscopic transit observations of the Neptunian ridge planet \planet obtained with ESPRESSO and MAROON-X to measure its projected stellar obliquity $\lambda$ using the RM effect. Contrary to a previous estimate that suggested a polar orbit from CARMENES data, we found that the planet has a $\lambda$ consistent with being aligned. Our best $\lambda$ estimates are obtained using the two ESPRESSO transits with a classical and a reloaded RM analyses, for which we measure $\lambda=-15^{+17}_{-16}\degr$ and $\lambda=-8^{+16}_{-16}\degr$, respectively. Both results are consistent with each other within 1~$\sigma$ and are not consistent with a polar orbit.

The CARMENES observations that led to the polar orbit estimate were modelled with an informative prior on an inaccurate \vsini estimate of $3.8\pm0.9$~\kms \citep{demangeon2018wasp}. Here, we derived new stellar parameters from our ESPRESSO observations, showing that the star appears as a very slow rotator with $\vsini\leq2$~\kms. Our modelling of the RM effect, both using the classical and the reloaded methods, also suggests a low \vsini value, of $\vsini=0.44\pm0.08$~\kms from the classical RM analysis and $\vsini=0.40\pm0.11$~\kms from the reloaded RM one. Together with telluric contamination and low-S/N in the CARMENES data, the inaccurate \vsini could be partly responsible for the polar obliquity measurement, since the RM effect can show degeneracies between $\lambda$ and \vsini. It is also possible that the CARMENES data did not have enough information to constrain the true RM signal, and that the polar orbit was due to systematics in the data. Our work highlights the importance of well-constrained stellar parameters in RM analyses.

We also re-analysed all \textit{TESS} transits, two new NGTS transits, and archival out-of-transit RVs to derive updated orbital parameters for the WASP-156 system.
To explore potential formation scenarios, we performed a population synthesis analysis to study the presence of planetary and brown dwarf companions in the residual RVs. We find that this allows us to rule out Jupiter-mass companions within $\sim$5~au in this system.
High-resolution imaging observations also rule out the presence of stellar companions within $\sim$500~au.

The observed close-in, nearly circular, and aligned orbit of \planet is consistent with early-on disc-driven migration. 
High-eccentricity migration mechanisms require a companion capable of exciting a strong eccentricity and might lead to a misaligned obliquity. The tidal realignment timescale necessary to re-align the planet into its present-day aligned orbit exceeds the age of the Universe. Therefore, the lack of nearby companions and the observed aligned orbit make high-eccentricity migration unlikely. Co-planar high-eccentricity migration, which preserves the obliquity, could still be possible if an eccentric, aligned massive Jupiter companion exists at more than $\sim10$~au.

Several works have suggested a tentative bi-modality in the obliquity distribution consistent in planets aligned or in polar orbits. With our new results, \planet moves from the tentative cluster of polar obliquities to the abundant group of planets with aligned orbits.
Regarding the short-period Neptunian landscape, recent works propose three different regimes: a desert of short period planets, a higher density of planets in a savanna at longer periods, and an overabundance of planets in a ridge in between. This variety in the population suggests different formation and evolution mechanisms, with potential observables in the density and obliquity distribution of the planets. \planet sits in the middle of the ridge, which shows a large diversity of obliquities and a tentative trend of decreasing density with increasing period, a potential sign of photoevaporation. Our work suggests that \planet arrived at its current position via disc-driven migration and survived photoevaporation. The number of Neptunian planets with known obliquities is still small, however, programmes such as ATREIDES \citep{bourrier2025atreides} or POSEIDON \citep{Espinoza-Retamal2026} are currently working towards increasing the sample and understanding the formation and migration pathways of the population of short-period Neptunes.


\section*{Acknowledgements}

We thank the referee for their thorough report, which helped improve the work presented in this article.
We thank Alex Polanski for kindly sharing the HIRES RVs of WASP-156, Steven Giacalone and Elise Furlan for kindly sharing information on the Shane and Zorro observations of WASP-156, and Andreas Seifahrt for sharing details about the MAROON-X observations.
We also thank Ares Osborn for help with analysis of the long-term ground-based photometric data.
This work is based on observations made with ESO Telescopes at the La Silla Paranal Observatory under the programme 109.23FU. Based on data obtained from the ESO Science Archive Facility with DOI: \url{https://doi.org/10.18727/archive/21}.
This work is also based on observations obtained at the international Gemini Observatory, a program of NSF NOIRLab, which is managed by the Association of Universities for Research in Astronomy (AURA) under a cooperative agreement with the U.S. National Science Foundation on behalf of the Gemini Observatory partnership: the U.S. National Science Foundation (United States), National Research Council (Canada), Agencia Nacional de Investigación y Desarrollo (Chile), Ministerio de Ciencia, Tecnología e Innovación (Argentina), Ministério da Ciência, Tecnologia, Inovações e Comunicações (Brazil), and Korea Astronomy and Space Science Institute (Republic of Korea). This work was enabled by observations made from the Gemini North telescope, located within the Maunakea Science Reserve and adjacent to the summit of Maunakea. We are grateful for the privilege of observing the Universe from a place that is unique in both its astronomical quality and its cultural significance.
The MAROON-X team acknowledges funding from the David and Lucile Packard Foundation, the Heising-Simons Foundation, the Gordon and Betty Moore Foundation, the Gemini Observatory, the NSF (award number 2108465), and NASA (grant number 80NSSC22K0117).
This paper made use of data collected by the \tess mission and are publicly available from the Mikulski Archive for Space Telescopes (MAST) operated by the Space Telescope Science Institute (STScI). Funding for the \tess mission is provided by NASA's Science Mission Directorate. We acknowledge the use of public \tess data from pipelines at the \tess Science Office and at the \tess Science Processing Operations Center. Resources supporting this work were provided by the NASA High-End Computing (HEC) Program through the NASA Advanced Supercomputing (NAS) Division at Ames Research Center for the production of the SPOC data products.
This work is based in part on data collected under the NGTS project at the ESO La Silla Paranal Observatory. The NGTS facility is operated by the consortium institutes with support from the UK Science and Technology Facilities Council (STFC) projects ST/M001962/1, ST/S002642/1 and ST/W003163/1.
This work has made use of data from the European Space Agency (ESA) mission {\it Gaia} (\url{https://www.cosmos.esa.int/gaia}), processed by the {\it Gaia}
Data Processing and Analysis Consortium (DPAC, \url{https://www.cosmos.esa.int/web/gaia/dpac/consortium}). Funding for the DPAC has been provided by national institutions, in particular the institutions participating in the {\it Gaia} Multilateral Agreement.
This research has made use of the Exoplanet Follow-up Observation Program (ExoFOP; DOI: 10.26134/ExoFOP5) website, which is operated by the California Institute of Technology, under contract with the National Aeronautics and Space Administration under the Exoplanet Exploration Program.
We acknowledge the use of the ExoAtmospheres database during the preparation of this work.
This paper makes use of data produced by the HATPI project, located in Chile at Las Campanas Observatory of the Carnegie Institution for Science and operated by the Department of Astrophysical Sciences at Princeton University. External funding for HATPI has been provided by the Gordon and Betty Moore Foundation, the David and Lucile Packard Foundation, the Mount Cuba Astronomical Foundation, and the Agencia Nacional de Investigación y Desarrollo (ANID) of Chile through QUIMAL, Millennium and Fondecyt grants.
This work is in part funded with a UKRI Future Leader Fellowship grant numbers MR/S035214/1 and MR/Y011759/1, UKRI grant EP/X027562/1, and STFC grant ST/X001121/1.
Computing facilities were provided by the Scientific Computing Research Technology Platform of the University of Warwick.
A.M. acknowledges funding from a UKRI Future Leader Fellowship, grant number MR/X033244/1 and a STFC small grant ST/Y002334/1.
A.V.F. acknowledges the support of the Institute of Physics through the Bell Burnell Graduate Scholarship Fund.
L.M. acknowledges support for this work by research funds from PRIN\,MUR\,2022 project 2022J4H55R.
J.S.J. gratefully acknowledges support by FONDECYT grant 1240738 and from the ANID BASAL project FB210003.
J.S. acknowledges support from STFC under grant number ST/Y002563/1.
J.V. acknowledges financial support from ANID / Fondo ALMA 2024 / No. 31240039, “On the origin of Warm-Giant Planets”.
V.K. acknowledges funding from the Royal Society through a Newton International Fellowship with grant number NIF/R1/232229.
X.L. acknowledges support from the Natural Science Foundation Youth Program of Sichuan Province (Grant No. 2024NSFSC1363) and the Doctoral Initiation Fund of China West Normal University (Grant No. 22kE036).
Together with the software mentioned in the main text, this work made use of \texttt{numpy} \citep{harris2020numpy}, \texttt{scipy} \citep{virtanen2020scipy}, \texttt{astropy} \citep{astropycollaboration2013astropy,astropycollaboration2018astropy}, and \texttt{matplotlib} \citep{hunter2007matplotlib}.

\section*{Data Availability}


The HARPS and ESPRESSO data used in this work are publicly available in the ESO archive under programmes 0102.C-0618(A) and 109.23FU. The MAROON-X data used in this work is publicly available in the Gemini archive under programme GN-2021B-Q-218. 
The reduced NGTS data used here is provided as online material. Raw NGTS data is available upon request to the authors and will be made available on the ESO archive on a future general release by NGTS.
The \tess data used in this work is publicly available in MAST.
The Zorro data used in this work is available in ExoFOP.
The ASAS-SN data used here is available online in the ASAS-SN archive. The HATPI data used here is available online in the HATPI archive.




\bibliographystyle{mnras}
\bibliography{rm_wasp156} 




\appendix

\section{Observational parameters}\label{sec:obs_fig}

\begin{figure}
\centering
\includegraphics[width=\columnwidth]{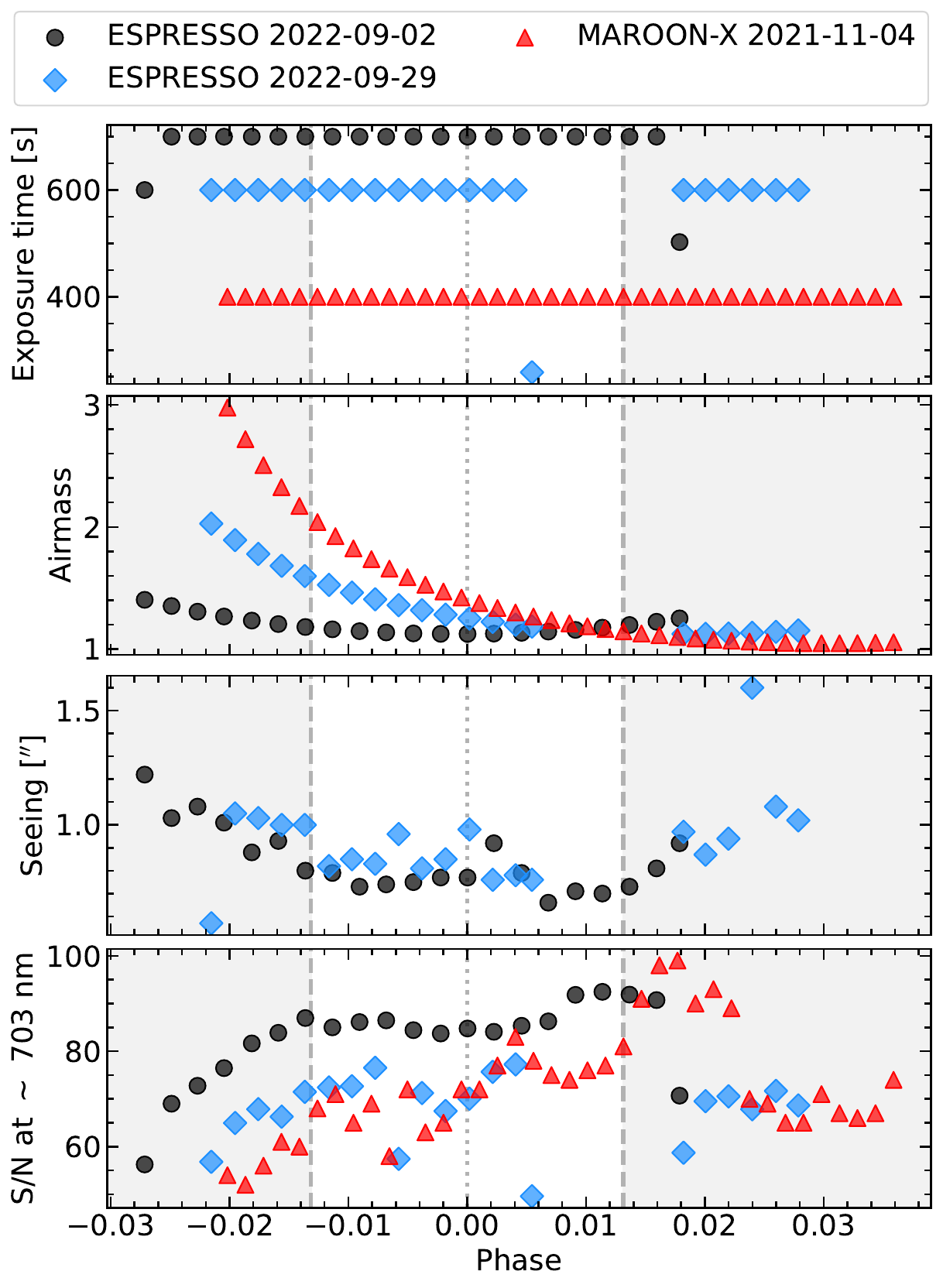}
\caption{From top to bottom: exposure time, airmass, seeing, and S/N for the ESPRESSO (first night in black circles, second night in blue diamonds) and MAROON-X (red triangles) observations as a function of the planet orbital phase. The seeing corresponds to the delivered seeing corrected by airmass. We show the median S/N of order 87 (central wavelength $\sim$703~nm), which is one of the orders with highest S/N for both spectrographs for WASP-156. Note that the S/N values shown are the sum in quadrature of the individual S/N values of each of the different slices (fibres) of order 87 (two for ESPRESSO and three for MAROON-X). Grey areas indicate out-of-transit phases, dashed lines indicate the transit start ($T_1$) and end ($T_4$), and dotted lines indicates mid-transit time ($T_0$).}
\label{fig:obs}
\end{figure}

\section{Stellar rotation period analysis}\label{sec:prot_appendix}

\begin{figure*}
\centering
\includegraphics[width=\textwidth]{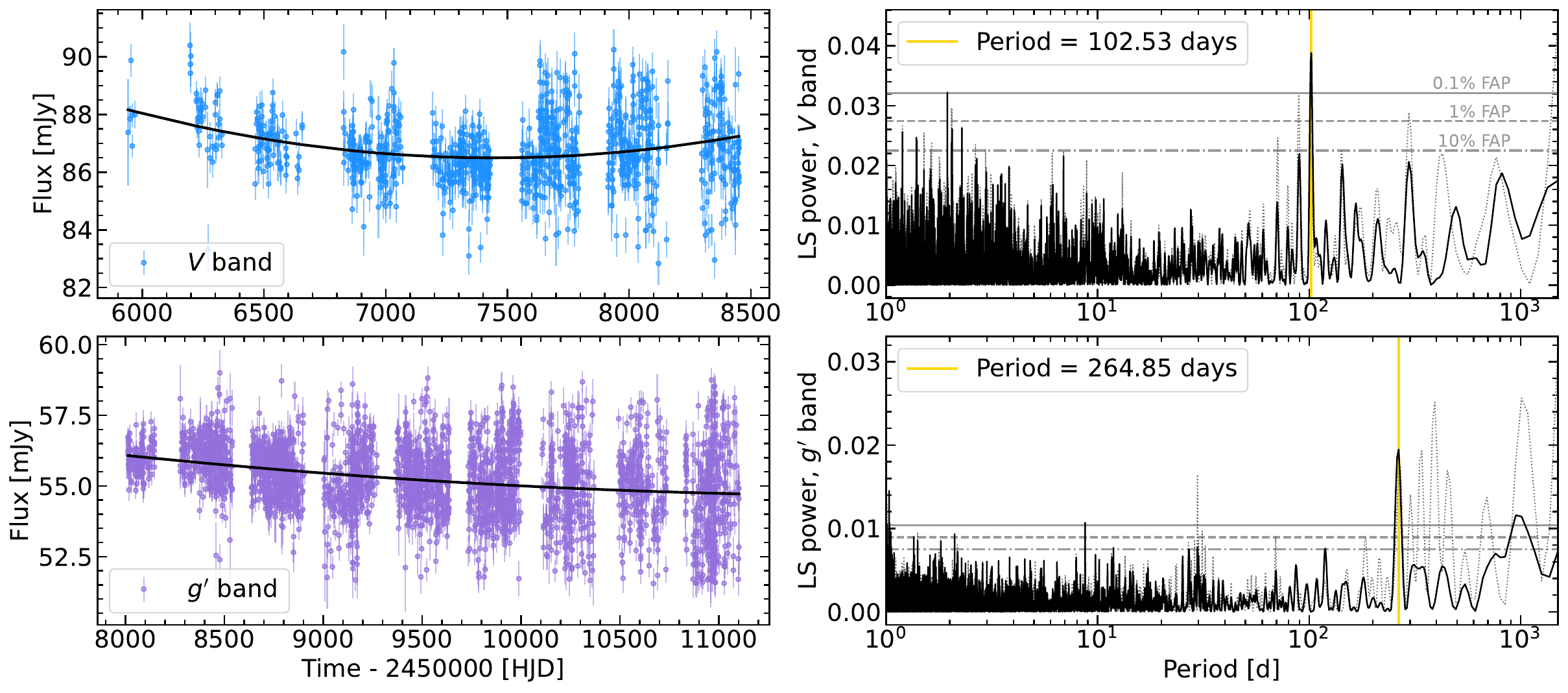}
\caption{ASAS-SN WASP-156 data. Left: Observations in the $V$ (top) and $g'$ (bottom) bands (circles), with the two-degree polynomial fit as the solid black line. Right: Generalised LS of the $V$ (top) and $g'$ (bottom) band datasets, where the solid black line corresponds to the LS computed after detrending the data by the polynomial, the dotted grey line shows the LS before detrending, and the solid, dashed, and dashed-dotted horizontal grey lines show the false alarm probability (FAP) levels for 0.1, 1, and 10\% FAP. The solid yellow vertical lines show the highest peak of the detrended periodogram.}
\label{fig:asasn}
\end{figure*}

\begin{figure*}
\centering
\includegraphics[width=\textwidth]{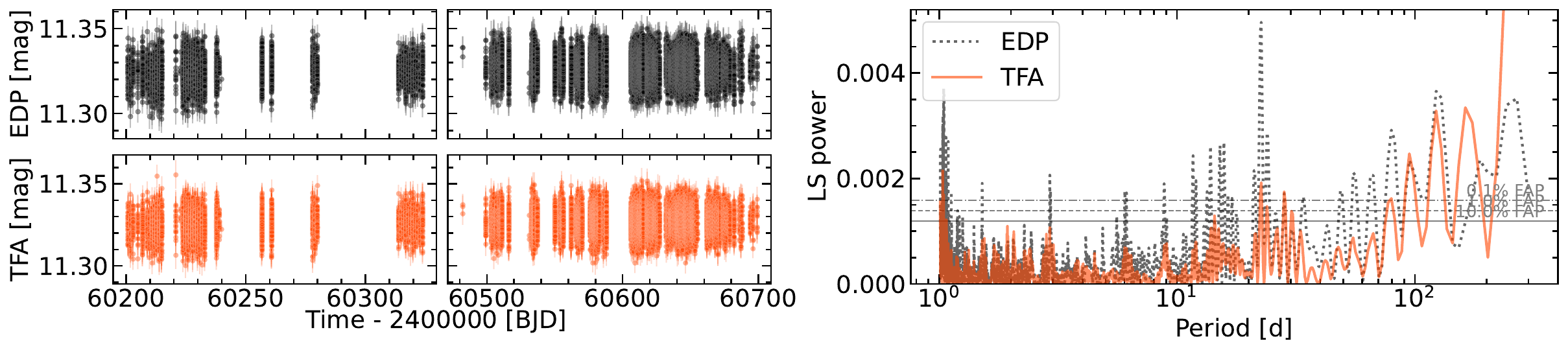}
\caption{HATPI WASP-156 data. Left: The two epochs of data, EDP-detrended (top, black circles) and TFA-detrended (bottom, orange circles). Right: Generalised LS of the EDP and TFA data (dotted black line and solid orange line, respectively). The solid, dashed, and dashed-dotted horizontal grey lines show the power corresponding to 0.1, 1, and 10\% FAP.}
\label{fig:hatpi}
\end{figure*}

We used archival WASP-156 data from ASAS-SN and HATPI to look for signals related to the stellar rotation.
We used the Sky Patrol portal to recompute ASAS-SN light curves. The data consists of 1035 observations in the $V$ band taken between 2012/01/15 and 2018/11/28, and 3604 observations in the $g'$ band taken between 2017/09/17 and 2026/03/02. We used the aperture photometry flux. We discarded observations with large uncertainties in the flux ($>2$~mJy, note that most of the observations had flux uncertainties well below 1~mJy, with a median of $\sim$0.4~mJy) and afterwards performed an iterative 3-$\sigma$ clipping two times. This process decreased the number of observations to 981 and 3336 for the $V$ and $g'$ band datasets, respectively. Finally, we detrended the light curves with a two-degree polynomial fit. We then computed the generalised Lomb-Scargle (LS) periodogram \citep{Zechmeister2009} of the data of each band both before and after detrending. Our frequency grid spanned from 1.01~d (to avoid strong 1~d signals) up to 1500~d (about half the span of the $g'$ band dataset). The cleaned data and corresponding periodograms can be seen in Fig. \ref{fig:asasn}.

Similarly, we also looked for periodicities in HATPI data. WASP-156 was observed by HATPI in two different seasons, between 2023/06/12 and 2024/02/03 (14711 observations), and between 2024/06/12 and 2025/02/03 (23995 observations). We used the aperture photometry flux, extracted using 1.95~pixels (index 0), following the recommendation of the HATPI database given the brightness of our target. The HATPI data is detrended via external parameter decorrelation (EPD, using e.g. airmass or image position), and further using a trend-filtering algorithm (TFA, which removes common systematics, applied after EPD). We looked for periodicities in both detrended data sets, using a frequency grid from 1.01 to 300~d (again about half the time span of the dataset). The EDP and TFA data, and their corresponding periodograms, can be seen in Fig. \ref{fig:hatpi}.

The ASAS-SN data (both before and after detrending) show several significant peaks at long periods. However, there is no common periodicity between the $V$ and the $g'$ band data. The HATPI data (both with the EDP and with the TFA detrending) also show significant peaks close to $\sim20$~d and at longer periods, but none coincide with the peaks observed in the ASAS-SN periodograms. We conclude that we cannot derive a clear estimate for the stellar rotation period from these data.

\section{Additional parameters from the \texttt{juliet} fit}\label{sec:juliet_extra}

\begin{table}
    \centering
    \caption{Summary statistics of the additional posteriors of the \texttt{juliet} fit, in addition to those presented in Table \ref{tab:juliet_params}.}
    \begin{tabular}{lr}
    \hline
       Sampled Parameter & Posterior \\ \hline
    \hline
       $\gamma_{\rm CORALIE}$ (m s$^{-1}$) & $9630.5\pm2.5$\\
       $\sigma_{\rm CORALIE}$ (m s$^{-1}$) & $0.1^{+1.6}_{-0.1}$\\
       $\gamma_{\rm SOPHIE}$ (m s$^{-1}$) & $9589\pm1$\\
       $\sigma_{\rm SOPHIE}$ (m s$^{-1}$) & $5.2\pm1.4$\\
       $\gamma_{\rm HARPS}$ (m s$^{-1}$) & $9650\pm2$\\   
       $\sigma_{\rm HARPS}$ (m s$^{-1}$) & $6.4^{+2.3}_{-1.5}$\\
       $\gamma_{\rm HIRES}$ (m s$^{-1}$) & $0.8\pm0.8$\\   
       $\sigma_{\rm HIRES}$ (m s$^{-1}$) & $3.0^{+0.8}_{-0.6}$\\
    \hline
       $\sigma_{\rm TESS}$ (ppm) & $1.4^{+8.5}_{-1.2}$\\
       $m_{\rm flux, TESS}$ & $0.002\pm0.004$\\
       GP$_{\sigma, {\rm TESS}}$ (ppm) & $0.0105^{+0.0008}_{-0.0004}$\\
       GP$_{\rho, {\rm TESS}}$ (d) & $14.4^{+1.2}_{-1.0}$ \\
    \hline
       $\sigma_{\rm NGTS11}$ (ppm) & $0.02^{+17.15}_{-0.02}$\\
       $m_{\rm flux, NGTS11}$ & $0.007\pm0.002$\\
       $\theta_{0, {\rm NGTS11}}$ & $0.0071\pm0.0015$ \\
       
       $\sigma_{\rm NGTS12}$ (ppm) & $0.02^{+19.84}_{-0.02}$\\
       $m_{\rm flux, NGTS12}$ & $0.006\pm0.002$\\
       $\theta_{0, {\rm NGTS12}}$ & $0.007\pm0.0015$ \\

       $\sigma_{\rm NGTS13}$ (ppm) & $0.03^{+20.77}_{-0.03}$\\
       $m_{\rm flux, NGTS13}$ & $-0.0003\pm0.0018$\\
       $\theta_{0, {\rm NGTS13}}$ & $0.001\pm0.001$ \\

       $\sigma_{\rm NGTS14}$ (ppm) & $0.02^{+18.14}_{-0.02}$\\
       $m_{\rm flux, NGTS14}$ & $-0.002\pm0.002$\\
       $\theta_{0, {\rm NGTS14}}$ & $0.0003\pm0.0015$ \\

       $\sigma_{\rm NGTS21}$ (ppm) & $0.01^{+11.44}_{-0.01}$\\
       $m_{\rm flux, NGTS21}$ & $-0.001\pm0.001$\\
       $\theta_{0, {\rm NGTS21}}$ & $0.001\pm0.001$ \\

       $\sigma_{\rm NGTS22}$ (ppm) & $0.03^{+18.61}_{-0.03}$\\
       $m_{\rm flux, NGTS22}$ & $-0.001\pm0.001$\\
       $\theta_{0, {\rm NGTS22}}$ & $0.001\pm0.001$ \\

       $\sigma_{\rm NGTS23}$ (ppm) & $0.02^{+14.85}_{-0.02}$\\
       $m_{\rm flux, NGTS23}$ & $0.003\pm0.001$\\
       $\theta_{0, {\rm NGTS23}}$ & $0.004\pm0.001$ \\

       $\sigma_{\rm NGTS24}$ (ppm) & $0.02^{+14.24}_{-0.02}$\\
       $m_{\rm flux, NGTS24}$ & $0.004\pm0.001$\\
       $\theta_{0, {\rm NGTS24}}$ & $0.005\pm0.001$ \\

       $\sigma_{\rm NGTS25}$ (ppm) & $0.02^{+14.06}_{-0.02}$\\
       $m_{\rm flux, NGTS25}$ & $0.003\pm0.001$\\
       $\theta_{0, {\rm NGTS25}}$ & $0.004\pm0.001$ \\
    \hline
    \end{tabular}
    \label{tab:juliet_params_extra}
\begin{flushleft}
{\it Notes:} For the RV instruments, $\gamma_\mathrm{inst}$ is an independent RV offset, and $\sigma_\mathrm{inst}$ is the jitter term. 
For \tess, $\sigma_{\rm TESS}$ is a jitter term added in quadrature to the nominal flux uncertainties, $m_{\rm flux, TESS}$ is the mean instrumental offset of the out-of-transit flux, GP$_{\sigma, {\rm TESS}}$ is the amplitude of the GP, and GP$_{\rho, {\rm TESS}}$ is the length-scale of the GP.
For the NGTS parameters, the first number in the subscript represents the night of observations (1 for 2022/09/02 and 2 for 2022/09/29), and the second number is an arbitrary number for each of the different cameras. Similarly to the \tess parameters, $\sigma_{\rm NGTSij}$ is a jitter term, $m_{\rm flux, NGTSij}$ if the out-of-transit flux offset, and $\theta_{0, {\rm NGTSij}}$ is the linear coefficient for the airmass detrending.
\end{flushleft}
\end{table}

\section{Classical RM additional figures}\label{sec:classicalrm_extra}

\begin{figure}
\centering
\includegraphics[width=\columnwidth]{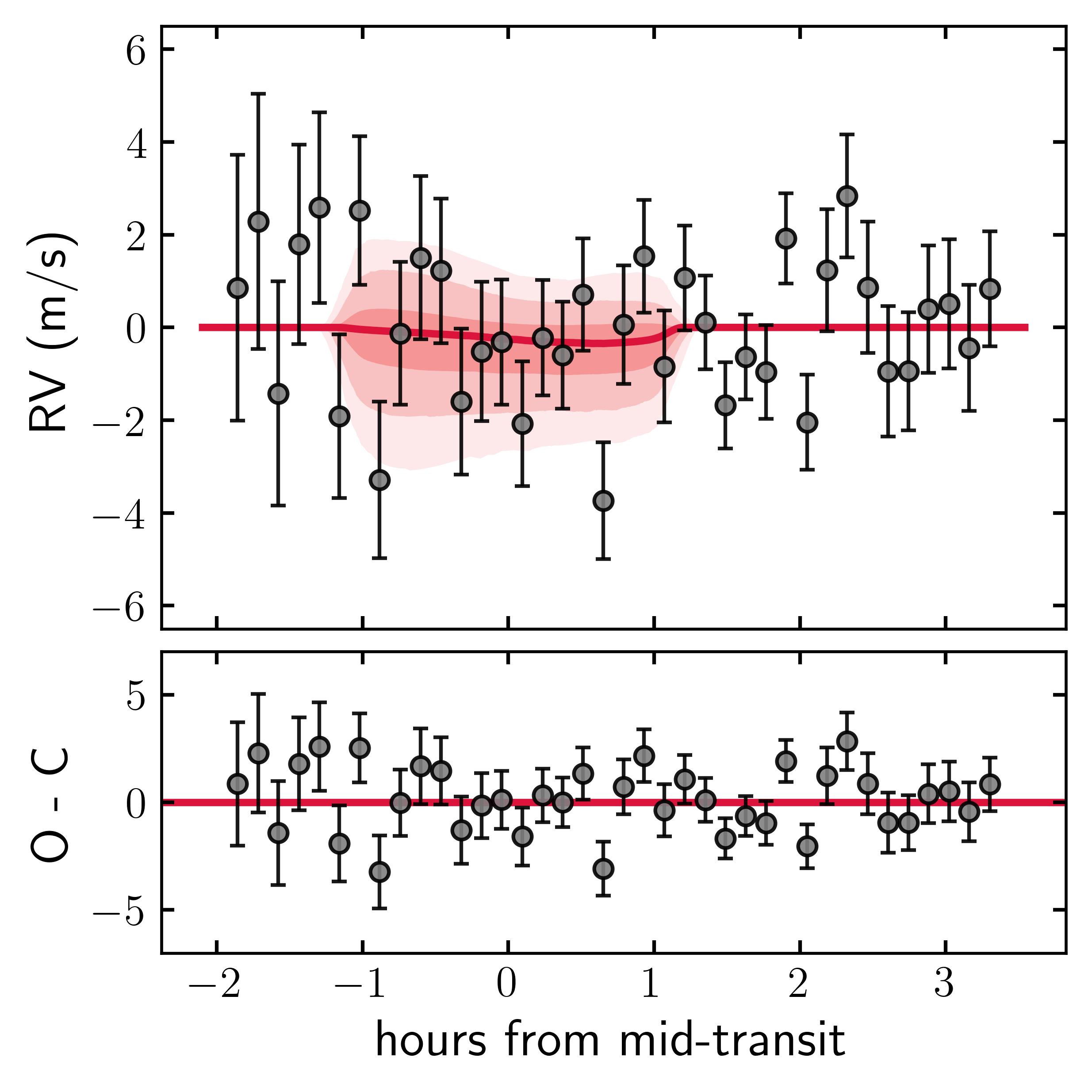}
\caption{MAROON-X \texttt{serval} disc-integrated RVs (grey circles, after subtracting the Keplerian motion induced by the planet) and best-fitting classical RM model (red line, with $1\,\sigma$, $2\,\sigma$, and $3\,\sigma$ intervals shown as the shaded regions).}
\label{fig:crm_fit_mx}
\end{figure}

\begin{figure*}
\centering
\includegraphics[width=0.33\textwidth]{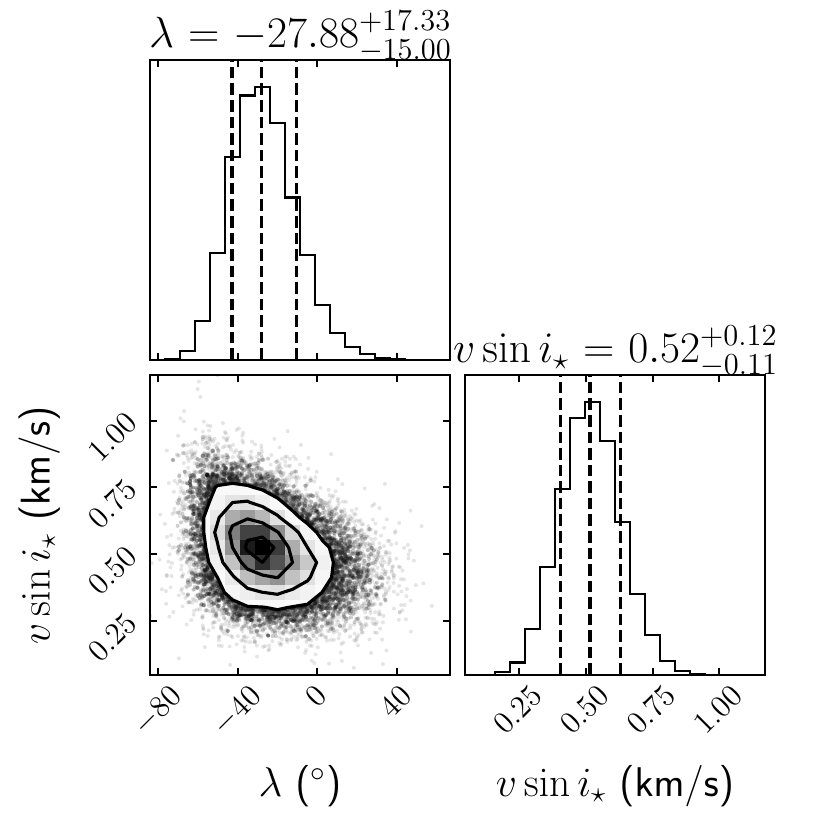}
\includegraphics[width=0.33\textwidth]{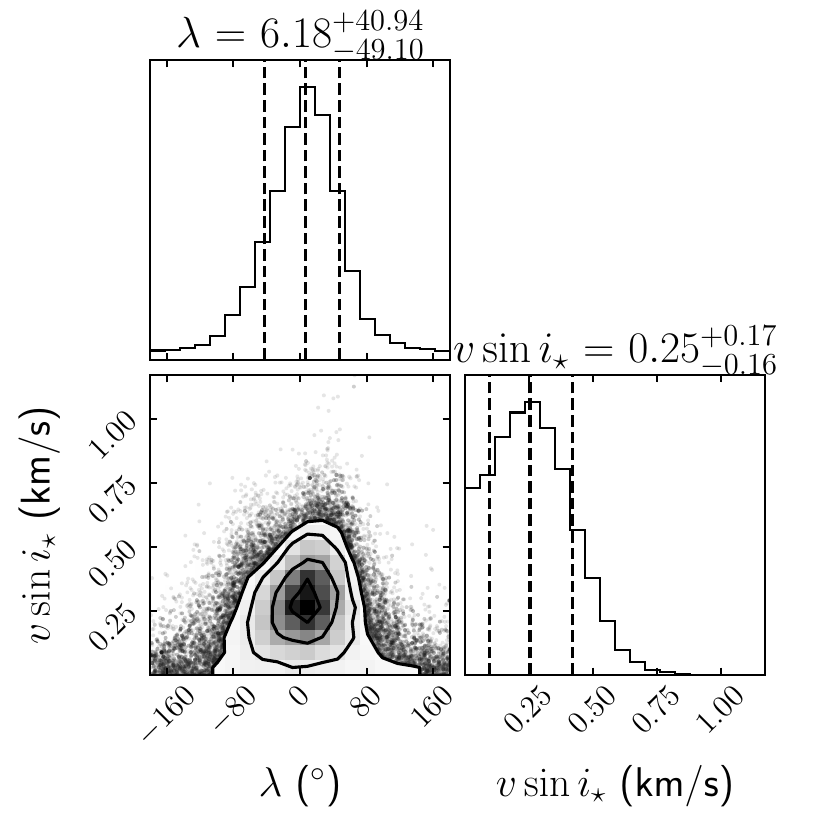}
\includegraphics[width=0.33\textwidth]{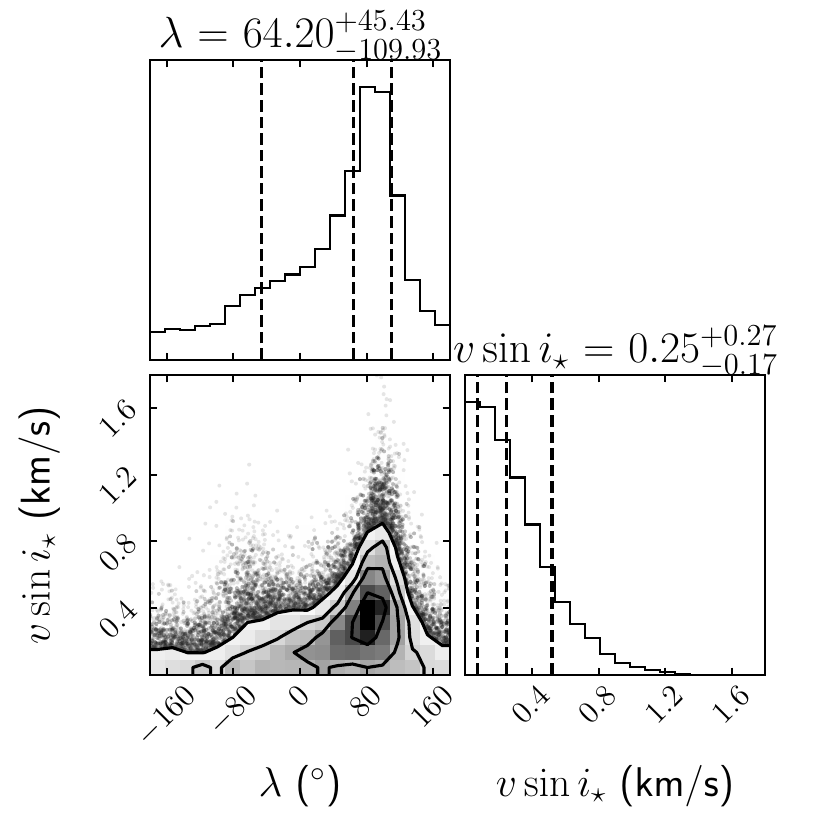}\\
\includegraphics[width=0.33\textwidth]{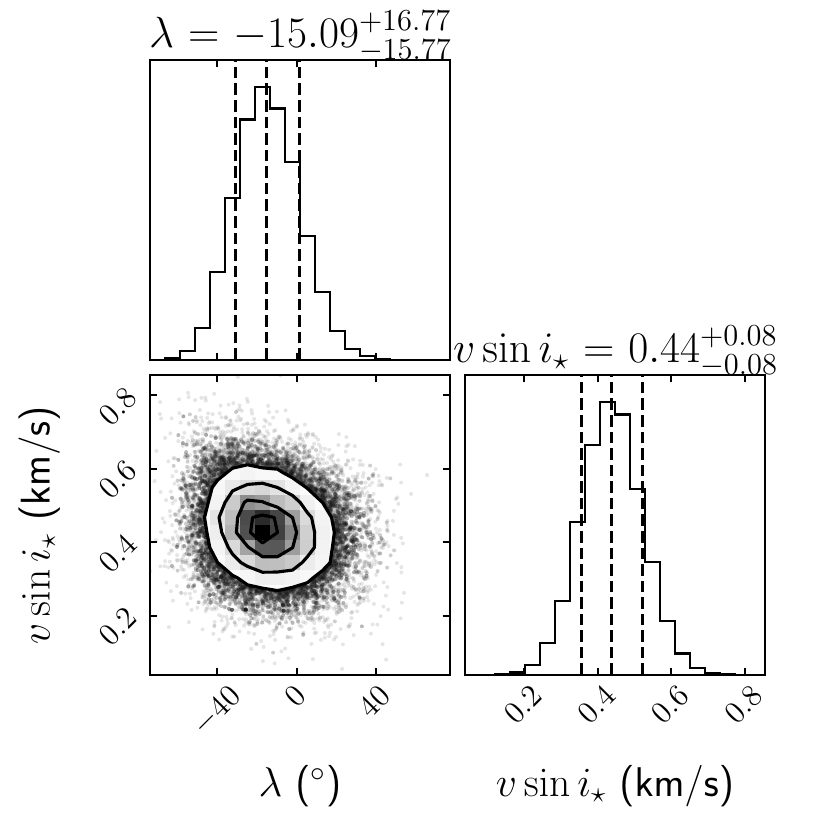}
\includegraphics[width=0.33\textwidth]{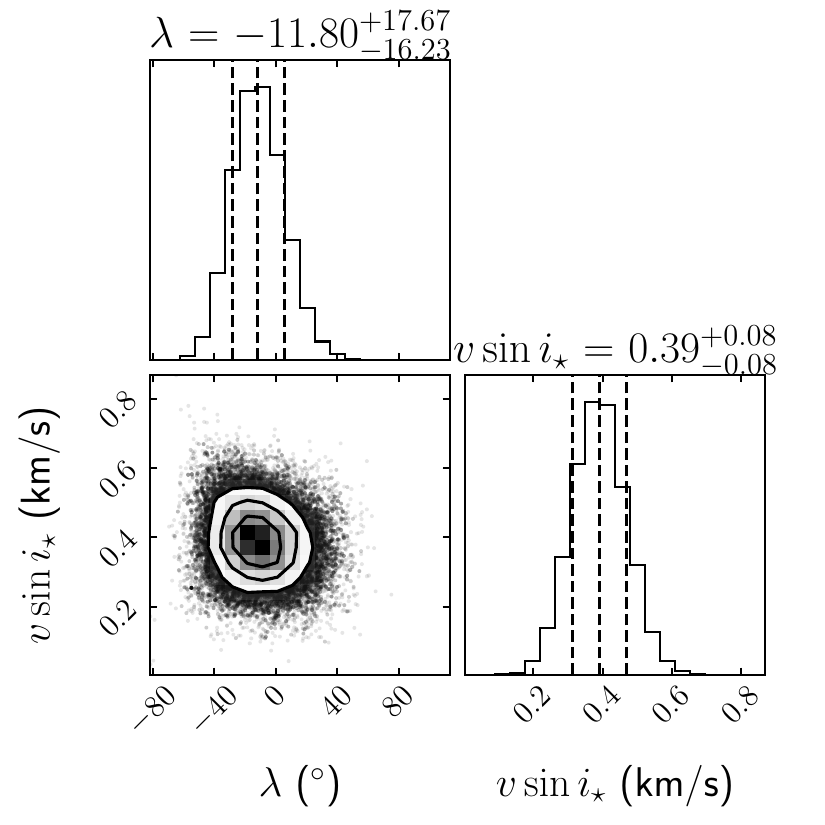}
\caption{Corner plot with the posteriors for the obliquity $\lambda$ and rotational velocity \vsini from the classical RM analysis of the first ESPRESSO night (top left), the second ESPRESSO night (top centre), the MAROON-X night (top right), both ESPRESSO transits (bottom left), and all three transits together (bottom right). We adopt the values from the fit to both ESPRESSO transits (bottom left).}
\label{fig:crm_corner_es}
\end{figure*}

\section{Reloaded RM additional table and figures}\label{sec:rrm_extra}

In addition to the SB models presented in Section \ref{sec:rrm_models}, we also considered more complex models that include, together with the SB rotation, centre-to-limb convective variations (CLV). 
The centre-to-limb convective variations are modelled with a first or second order polynomial radially symmetric from the stellar disc centre, where $c_1$ and $c_2$ are the first and second order coefficients, respectively. Again, for more details about the models, we refer the reader to \citet{cegla2016rrm}.

For these SB+CLV models we used the same MCMC method to fit the data as described in Section \ref{sec:rrm_models}. We used uninformative priors (i.e., unconstrained) for $c_1$ and $c_2$ with initial values for our walkers of $c_1$ and $c_2$ $=0$~\kms. Different initial values for these parameters do not affect the final results of the fit.

For each set of observations (individual transits, both ESPRESSO transits, and all three transits together) we fit a SB rotation in conjunction with a linear CLV, and SB rotation with quadratic CLV. Note that we also preliminarily considered more complex models that account for differential rotation, but discarded them due to our data not being sufficiently precise to see such effects. As done for the SB model, we also considered a reduced set of observations with  $\langle\mu\rangle<0.3$. We also found that the results obtained from this reduced data set agree well within $1\,\sigma$ with those using the full data set, and hence do not consider them further.
 
We assessed the goodness of fit with the reduced chi-squared statistic ($\chi^2_\mathrm{red}$) and the Bayesian information criterion (BIC). We note that the BIC values can only be compared for models fitted to the same dataset.
 Our full results for the SB and CLV models are shown in Table \ref{tab:results_rrm_all}.

The SB+CLV models, which are more complex than the SB model, tend to overfit the data ($\chi^2_\mathrm{red}<1$) and/or do not show a significant enough difference compared to the simpler SB models \citep[the difference in BIC is less than $\simeq6$, which is the minimum difference for a model to be considered a better fit,][]{lorah2019bic,raftery1995bic}.

\begin{table*}
\centering
\caption{RRM fit results for the different sets of observations considered and different models tested.}
\begin{tabular}{lcccccccc}
\hline
Model & $\lambda$& \vsini & $c_1$ &$c_2$ & \# in-transit & \# model& $\chi^2_\mathrm{red}$ & BIC \\
      & (\degr)  & (\kms) & (\kms)&(\kms)&obs.           & param. &                       &     \\ \hline
\multicolumn{8}{l}{\textbf{ESPRESSO 2022-09-02}} \\
SB, $\langle\mu\rangle\geq0.0$      & $7^{+17}_{-17}$  & $0.47^{+0.12}_{-0.12}$ & $-$ & $-$ & 11 & 2 & 1.10 & 14.65 \\[0.5ex] 
SB CLV1, $\langle\mu\rangle\geq0.0$ & $25^{+18}_{-17}$ & $0.49^{+0.13}_{-0.12}$ & $-0.85^{+0.36}_{-0.36}$ & $-$ & 11 & 3 & 0.50 & 11.11 \\[0.5ex] 
SB CLV2, $\langle\mu\rangle\geq0.0$ & $34^{+17}_{-15}$ & $0.55^{+0.13}_{-0.13}$ & $-4.9^{+2.6}_{-2.6}$ & $3.0^{+1.9}_{-2.0}$ & 11 & 4 & 0.17 & 10.78 \\[0.5ex] 
\multicolumn{8}{l}{\textbf{ESPRESSO 2022-09-29}} \\
SB, $\langle\mu\rangle\geq0.0$      & $-74^{+34}_{-45}$ & $0.35^{+0.20}_{-0.20}$ & $-$ & $-$ & 10 & 2 & 1.93 & 20.01\\[0.5ex] 
SB CLV1, $\langle\mu\rangle\geq0.0$ & $-112^{+8}_{-12}$ & $1.32^{+0.48}_{-0.47}$ & $2.35^{+0.96}_{-0.93}$ & $-$ & 10 & 3 & 0.99 & 13.80 \\[0.5ex] 
SB CLV2, $\langle\mu\rangle\geq0.0$ & $-108^{+9}_{-13}$ & $1.33^{+0.45}_{-0.46}$ & $7.7^{+4.4}_{-4.3}$ &$-4.1^{+3.3}_{-3.2}$ & 10 & 4 & 0.89 & 14.57 \\[0.5ex] 
\multicolumn{8}{l}{\textbf{ESPRESSO 2022-09-02 \& 2022-09-29}} \\[0.5ex] 
SB, $\langle\mu\rangle\geq0.0$      &  $-8^{+16}_{-16}$ & $0.40^{+0.11}_{-0.11}$ & $-$ & $-$ & 21 & 2 & 1.57 & 35.97 \\[0.5ex] 
SB CLV1, $\langle\mu\rangle\geq0.0$ &   $9^{+18}_{-17}$ & $0.42^{+0.11}_{-0.11}$ & $-0.56^{+0.31}_{-0.31}$ & $-$ & 21 & 3 & 1.47 & 35.54 \\[0.5ex] 
SB CLV2, $\langle\mu\rangle\geq0.0$ & $11^{+19}_{-17}$ & $0.43^{+0.11}_{-0.11}$ & $-1.5^{+2.2}_{-2.2}$ & $0.7^{+1.7}_{-1.7}$ & 21 & 4 & 1.54 & 38.39 \\[0.5ex] 
\multicolumn{8}{l}{\textbf{MAROON-X 2021-11-04}} \\
SB, $\langle\mu\rangle\geq0.0$      & $31^{+101}_{-77}$ & $0.13^{+0.09}_{-0.13}$ & $-$ & $-$ & 17 & 2 & 1.31 & 25.33 \\[0.5ex] 
SB CLV1, $\langle\mu\rangle\geq0.0$ & $31^{+102}_{-73}$ & $0.13^{+0.09}_{-0.14}$ & $-0.19^{+0.43}_{-0.42}$ & $-$& 17 & 3 & 1.39 & 27.90\\[0.5ex] 
SB CLV2, $\langle\mu\rangle\geq0.0$ & $22^{+113}_{-86}$ & $0.12^{+0.09}_{-0.14}$ & $1.8^{+2.6}_{-2.5}$ & $-1.6^{+2.0}_{-2.0}$ & 17 & 4 & 1.46 & 30.36 \\[0.5ex] 
\multicolumn{8}{l}{\textbf{ESPRESSO 2022-09-02 \& 2022-09-29, \& MAROON-X 2021-11-04}} \\
SB, $\langle\mu\rangle\geq0.0$      & $10^{+23}_{-21}$ & $0.260^{+0.098}_{-0.097}$ & $-$ & $-$ & 38 & 2 & 1.62 & 65.80 \\[0.5ex] 
SB CLV1, $\langle\mu\rangle\geq0.0$ & $8^{+20}_{-19}$ & $0.310^{+0.090}_{-0.089}$ & $-0.46^{+0.24}_{-0.25}$ & $-$ & 38 & 3 & 1.41 & 60.42 \\[0.5ex] 
SB CLV2, $\langle\mu\rangle\geq0.0$ & $7^{+20}_{-20}$ & $0.309^{+0.089}_{-0.087}$ & $-0.4^{+1.7}_{-1.7}$ & $0.0^{+1.3}_{-1.3}$ & 38 & 4 & 1.46 & 64.07 \\[0.5ex] 
\hline
\end{tabular}

\begin{flushleft}
{\it Notes}: We show the preferred models fitted to each transit individually, to both ESPRESSO transits together, and to all three transits together. Models: SB means solid body model, CLV1 means centre-to-limb linear model (polynomial of order 1, for which we fit $c_1$) and CLV2 means centre-to-limb quadratic model (polynomial of order 2, for which we fit $c_1$ and $c_2$).
\end{flushleft}
\label{tab:results_rrm_all}
\end{table*}

\begin{figure*}
\centering
\includegraphics[width=0.33\linewidth]{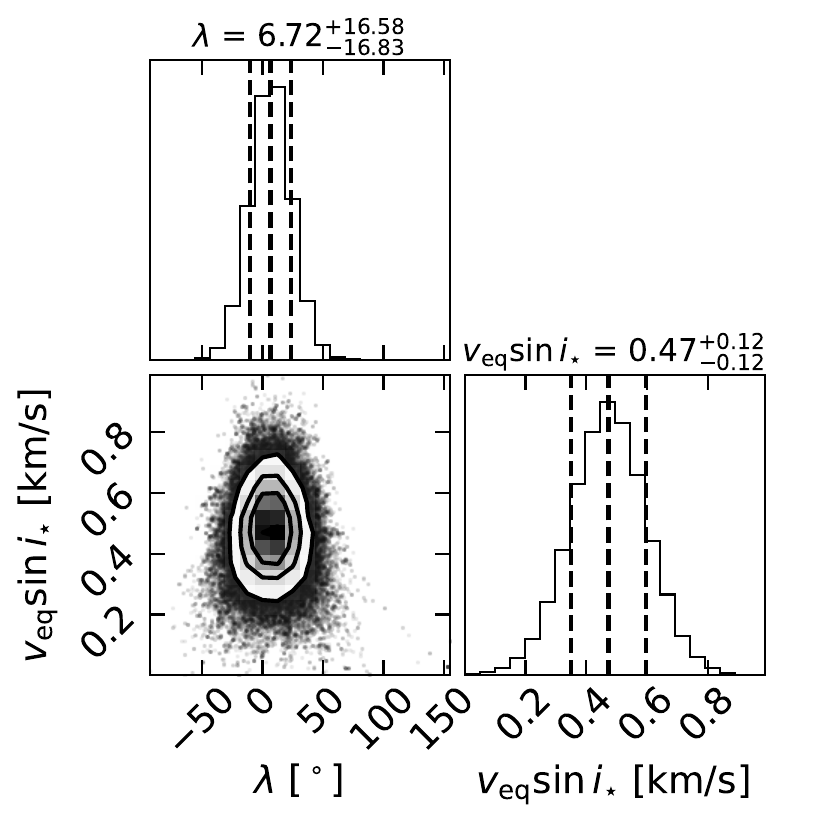}
\includegraphics[width=0.33\linewidth]{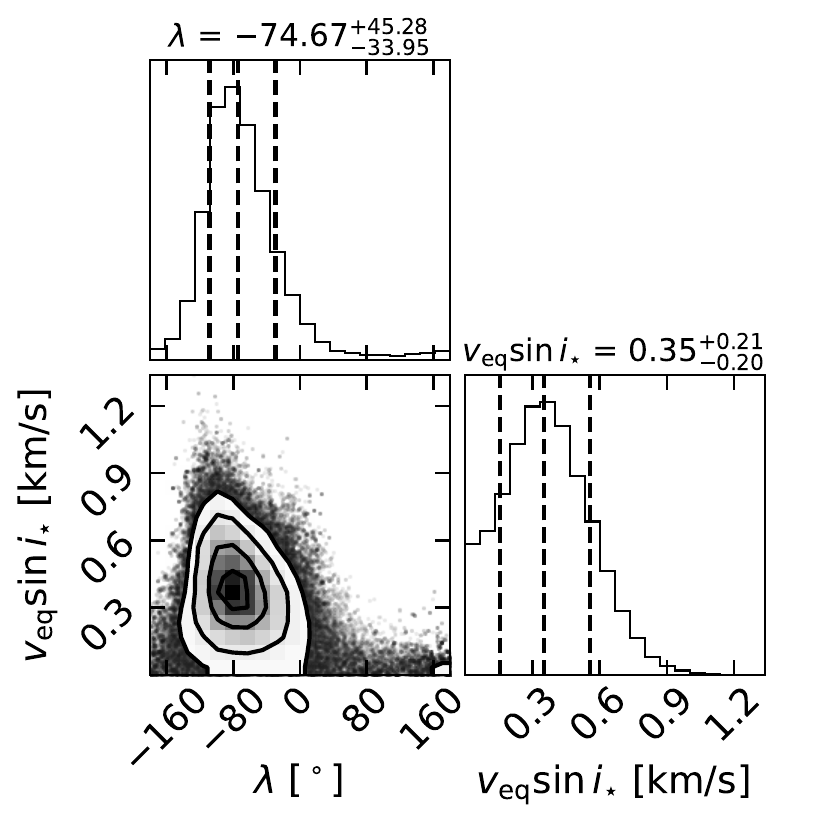}
\includegraphics[width=0.33\linewidth]{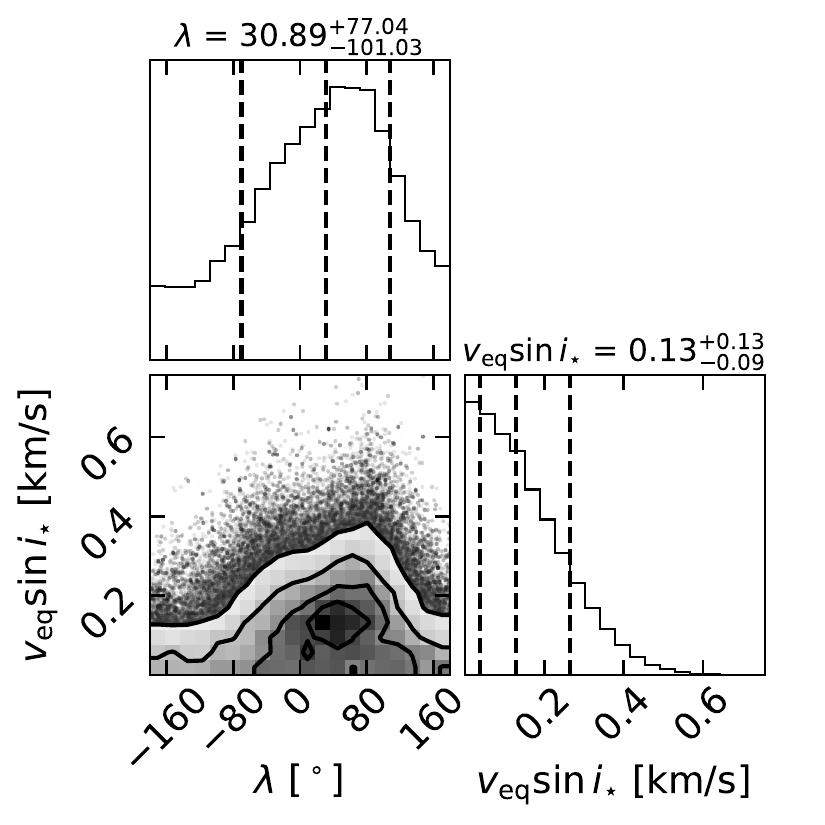}\\
\includegraphics[width=0.33\linewidth]{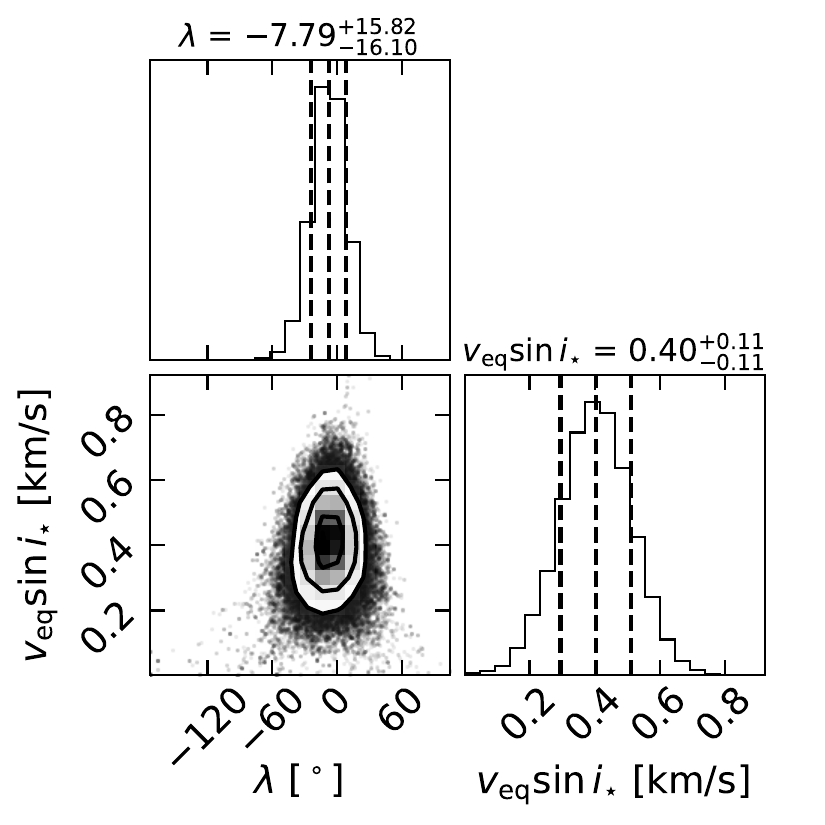}
\includegraphics[width=0.33\linewidth]{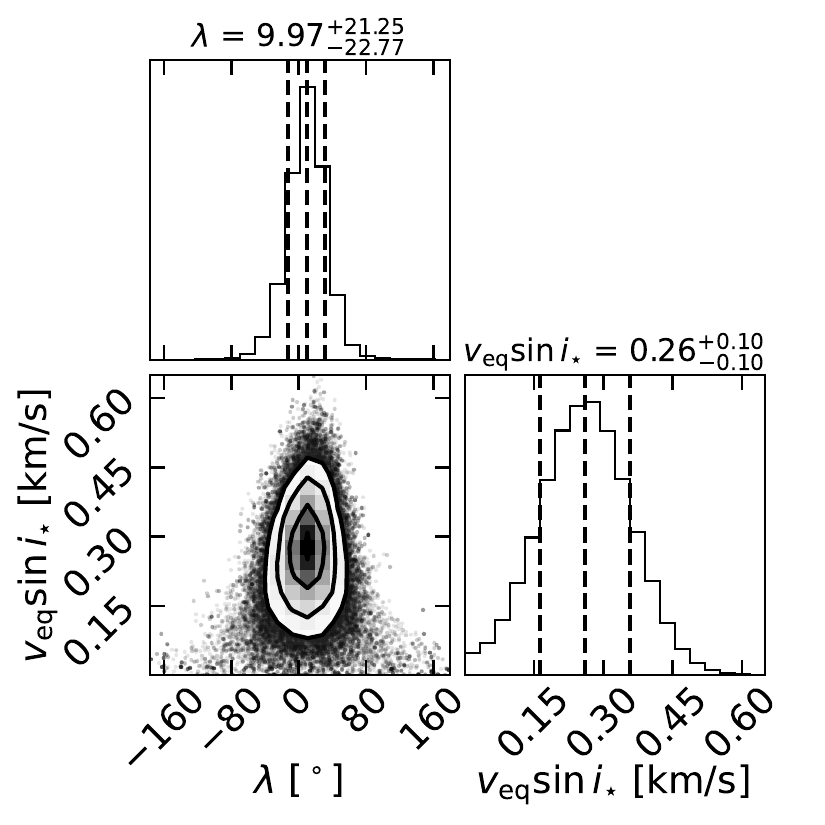}
\caption{Corner plot with the posteriors of the SB model fit from the reloaded RM analysis of the first ESPRESSO transit (top left), the second ESPRESSO transit (top centre), MAROON-X (top right), both ESPRESSO transits (bottom left), and all three transits (bottom right). We adopt the values from the fit to both ESPRESSO transits (bottom left).}
\label{fig:rrm_corner_sb_esmxall}
\end{figure*}

\section{3D obliquity distribution}

\begin{figure}
\centering
\includegraphics[width=\linewidth]{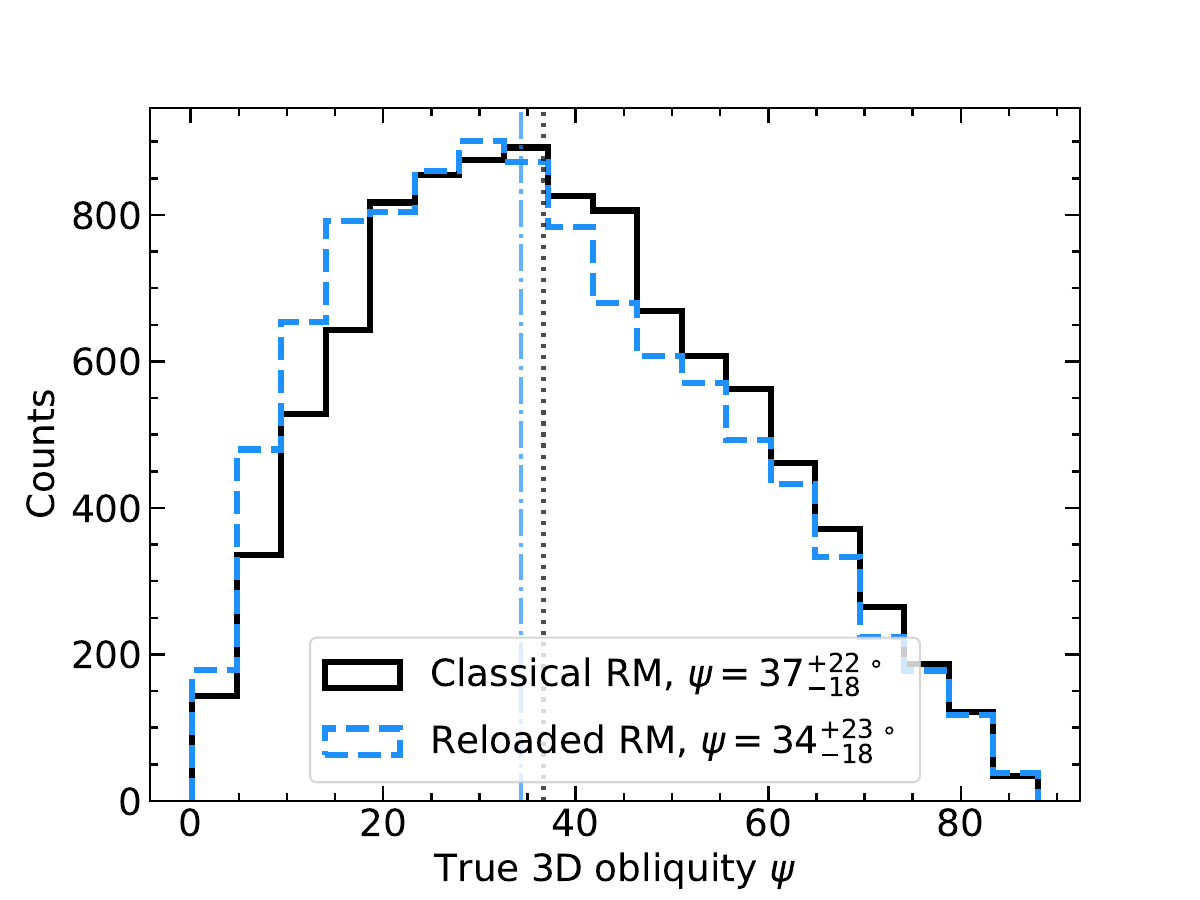}
\caption{3D obliquity distribution $\psi$ obtained after simulating random system configurations (see Sect. \ref{sec:discusionvalues}). The black solid-line histogram shows the distribution obtained using the $\lambda$ value from the classical RM, and the blue dashed-line histogram shows the same but for the reloaded RM $\lambda$. The vertical black dotted line and the blue dash-dotted line show the median values of each distribution, also shown in the legend.}
\label{fig:psi}
\end{figure}

\section{ExoFOP follow-up figures}\label{sec:exofop}

\begin{figure}
\centering
\includegraphics[width=\columnwidth]{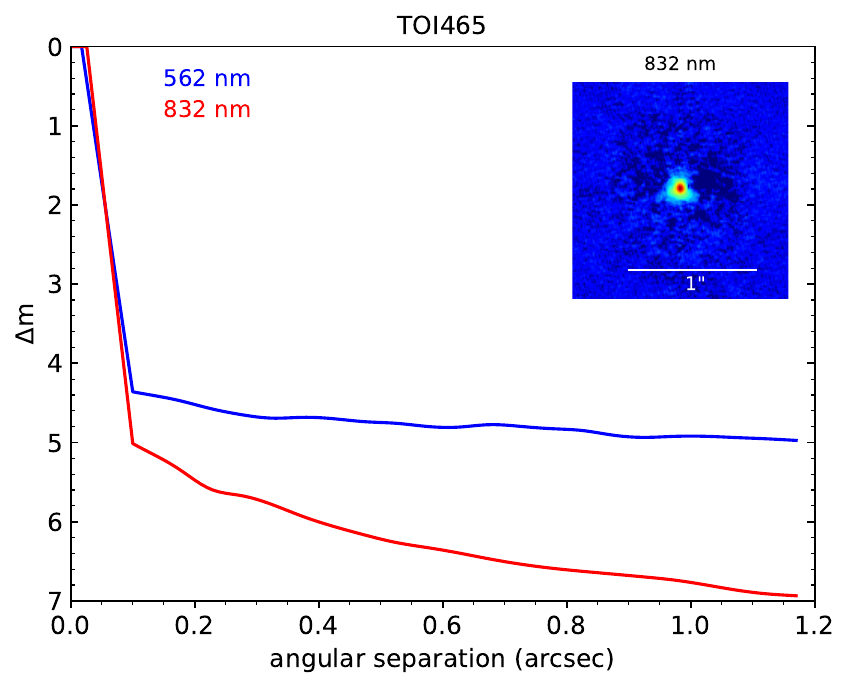}
\caption{Speckle imaging contrast curves of WASP-156 from the 562~nm and 832~nm bandpasses of the Zorro speckle imager at Gemini South Observatory, from the ExoFOP archive (PI: Howell). The inset shows the reconstructed image from the 832~nm filter. 
}
\label{fig:zorro}
\end{figure}


\bsp	
\label{lastpage}
\end{document}